\documentclass[10pt,twocolumn,amssymb,floats,prl,aps]{revtex4-1}
\usepackage[latin9]{inputenc}
\usepackage{amsmath}
\usepackage{amssymb}
\usepackage{graphicx}

\makeatletter

\usepackage{array}\usepackage[normalem]{ulem}
\usepackage{amsmath}
\usepackage{epsf}
\usepackage{dcolumn}
\usepackage{bm}

\usepackage{amssymb}
\usepackage{array}
\usepackage{varioref}
\usepackage{wrapfig}
\usepackage{etoolbox}
\usepackage{color}

\providecommand{\pd}{\partial}

\providecommand{\bv}[1]{\bm{\mathrm{#1}}}

\providecommand{\w}{\omega}
\providecommand{\W}{\Omega}
\providecommand{\Wp}{\W_{\rm ph}}
\providecommand{\q}{\bv{q}}

\renewcommand{\k}{\bv{k}}

\providecommand{\Tr}{\mbox{Tr}}

\renewcommand{\q}{\bv{q}}

\providecommand{\idx}{\int_{\mathcal{C}} dt \sum_{\x_{i}}}

\makeatother

\begin{document}

\title{Laser-induced
  control of an electronic nematic quantum phase
transition}

\author{Avraham Klein}

\author{Morten H. Christensen}

\author{Rafael M. Fernandes}

\affiliation{School of Physics and Astronomy, University of Minnesota, Minneapolis, MN
55455}
\begin{abstract}
Ultrafast techniques have emerged as promising methods to study and control quantum materials. To maintain the quantum nature of the systems under study, excess heating must be avoided. In this work, we demonstrate a method that employs the nonequilibrium  laser  excitation of planar stretching optical phonons in tetragonal systems to quench an electronic nematic state across a quantum phase transition. Appropriately tuned off-resonant pulses can perform a quantum quench of the system either into the nematic phase (red detuning) or out of it (blue detuning). The nonlinear coupling of this phonon mode to nematicity not only mediates interactions in the nematic channel, but it also suppresses heating effects. We illustrate the applicability of our general results by considering the microscopic parameters of the nematic unconventional superconductor FeSe.
\end{abstract}

\maketitle
\providecommand{\ph}{\mbox{ph}} \providecommand{\x}{\bv{x}} \providecommand{\X}{\bv{X}}
\global\long\def\AA{\textrm{A\kern-1.3ex  \raisebox{0.6ex}{\ensuremath{^{\circ}}}}}

\label{sec:introduction}

Nonequilibrium studies of strongly-correlated electron systems have undergone significant progress in recent years, due to both
theoretical advances~\cite{Mitra2016,F.2017,Heyl2018} and outstanding developments in ultrafast pump-probe techniques~\cite{Orenstein2012,Giannetti2016,Mitrano2016,Hoegen2018}. Light-based control of quantum materials has wide applications ranging from high-temperature superconductivity~\cite{Giannetti2016,Mitrano2016,Demler2016} to quantum computation~\cite{Sentef2018}. The basis for such control derives from the use of laser pulses to switch between different electronic states inhabiting the phase diagrams of these exotic materials.
A natural route to realizing such control is to exploit the electron-phonon coupling and precisely excite infrared-active optical phonon modes \cite{Rini2007,Nova2016,Gandolfi2017,Puviani2018,Hoegen2018} via strong terahertz laser pulses \cite{Hafez2016,Fulop19-26Aug.2017,Hoffmann2011,Nicoletti2016}.
Because infrared phonon modes do not couple directly to the electronic
charge density, the changes in electronic
properties
are mediated by nonlinear effects \cite{Georges2014,Kennes2017,Aleiner2018}.
Importantly, the ordered states that can be most efficiently controlled
by this approach are those that couple strongly to the lattice, such
as superconductivity or metal-to-insulator transitions \cite{Georges2014,Demler2016,Werner2017,Kennes2017}.
Identifying a method to optically control a strongly-correlated electronic phase, while maintaining its quantum, low temperature properties, would provide a key ingredient for manipulating the intertwined orders characteristic of quantum materials.

Electronic nematicity is intimately coupled to the lattice~\cite{Fernandes2010,Karahasanovic2016,Paul2017}.
When the electronic system spontaneously breaks rotational symmetry \cite{Kivelson1998}, it inevitably triggers a structural distortion. Nematic phases have been widely observed in quantum materials, from unconventional superconductors
such as cuprates, pnictides, and heavy fermions, to ruthenates and
semiconductors displaying the quantum Hall effect~
\cite{Fradkin2010,Vojta20092,Fernandes2014,Ronning2017}.
The driving mechanism of nematicity in these compounds and its relationship
with other phenomena such as superconductivity, magnetism, and charge-order,
are all hotly debated~\cite{Fernandes2014,Boehmer2017,Lederer2017}.
Most suggestively, in several of these materials it is widely believed
that the nematic phase ends at a putative quantum critical point (QCP)
\cite{Fisher2016}, which may host an exotic non-Fermi-liquid phase and a superconducting dome \cite{Hussey2019}.
Controlling the nematic degrees of freedom would
therefore provide
a high-precision tool for studying the competing and intertwined orders
in these systems. The strong coupling to the lattice \cite{Miller2015} suggests light control as a feasible route,
which has hitherto remained largely unexplored.

\begin{figure}
\includegraphics[width=\columnwidth]{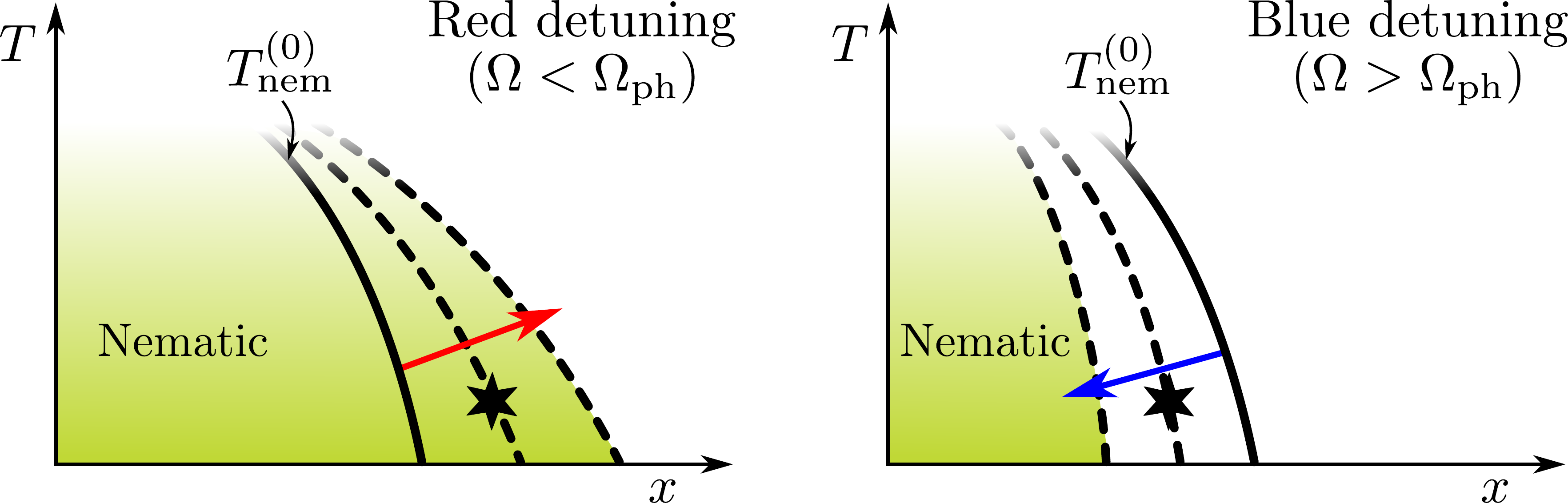}
\caption{\label{fig:schematic_phase_diagram}Schematic illustration of the proposed experimental setup. Depending on the detuning between the applied optical pulse and the $E_{u}$ phonon frequency, $\Omega_{\rm ph}$, the nematic phase can be either enhanced or suppressed. The solid line denotes the equilibrium nematic phase transition at $T^{(0)}_{\rm nem}$, while the dashed lines illustrate the non-equilibrium shift as the laser intensity is increased. The stars denote compositions that would undergo a nonequilibrium nematic transition.}
\end{figure}

In this Letter we develop a theory of nonequilibrium optical control
of a generic nematic phase on the square lattice, based on the off-resonance excitation of a particular infrared optical phonon mode.
Being entirely based on symmetry arguments, our model is independent of microscopic considerations, and is thus applicable to a wide variety of nematic materials. In particular, we demonstrate how our results offer an avenue to implement
a laser controlled quench across the nematic quantum phase transition, as illustrated schematically in Fig.~\ref{fig:schematic_phase_diagram}. This can be used to either promote nematic order in an otherwise disordered state, or to kill nematic order to enhance other competing states, such as superconductivity~\cite{Nandi2010}. We illustrate the applicability of our model by considering microscopic parameters of the unconventional superconductor FeSe, a posterchild of electronic nematic order.

The key ingredient in our analysis is that the onset of nematic
order breaks the tetragonal symmetry either by making the horizontal
and vertical bonds inequivalent (called $d_{x^{2}-y^{2}}$-wave, or
$B_{1g}$, nematic order) or the diagonals inequivalent (called $d_{xy}$-wave,
or $B_{2g}$, nematic order). For concreteness, we consider hereafter
the $B_{1g}$ case.
The orthorhombic lattice distortion that accompanies
nematic order is due to the linear coupling between the electrons
and a transverse acoustic phonon mode that propagates along the $[110]$
direction. The phonon velocity is strongly renormalized by nematic
fluctuations, and vanishes at the nematic transition. Although this
linear nemato-elastic coupling has been widely employed to investigate
the nematic properties of many materials~\cite{Fisher2016,Ronning2017},
the acoustic nature of the phonon makes it inconvenient for optical
control~\cite{Schuett2018}. Instead, we propose
to control the nematic phase by optically exciting the $E_{u}$ optical
phonon mode, corresponding to two degenerate planar stretching lattice
vibrations.

This infrared-active mode, ideal for laser manipulation,
is present in any tetragonal system. While symmetry forbids incoming photons and the $E_{u}$ phonon from coupling linearly to the electronic density, the $E_{u}$ vibrations do couple \emph{quadratically} to the electronic nematic degrees of freedom. Therefore, in contrast to traditional pump-probe setups, the lattice is excited directly, whereas the electronic nematic channel is only excited indirectly, via the nonlinear phonon coupling. As a result, one expects heating effects, which commonly plague the interpretation of pump-probe data, to be significantly reduced. Our proposal is thus similar in spirit to that of Ref.~\onlinecite{Kennes2017}, which proposed that the excitation of infrared-active phonons leads to an attractive electronic interaction that favors superconductivity, via the nonlinear coupling to the charge density. 

The differences between a quadratic \emph{versus} a linear coupling also play a crucial role in determining how a nematic phase transition can be controlled by externally-applied lattice perturbations. Physically, the linear coupling between electrons and the acoustic phonon is manifested by a change in the rest positions of the atoms of the square lattice once nematic order onsets. As a result, lattice displacements induced by external strain act as a conjugate field to the nematic order parameter. Thus, although strain can be used to assess the nematic susceptibility, it smears the nematic transition completely, making it very difficult to manipulate a nematic transition with strain. In contrast, the quadratic coupling to the optical phonon is manifested as a splitting of the resonance frequencies
of the degenerate $E_u$ phonon modes~\cite{Kennes2017,Sentef2017} in the presence of nematic order. Consequently, the external excitation of this phonon mode shifts the frequency of the low-energy collective nematic mode, without breaking explicitly the tetragonal symmetry of the system.
As we will show, the amplitude and phase of this shift can be controlled in a precise manner by an external optical pulse, and thus used to tune the system both towards and away from the nematic transition, in proportion to the intensity of the exciting laser.
The experimental consequences are shown in Fig.~\ref{fig:schematic_phase_diagram}. For laser frequencies below the phonon resonance frequency, $\Omega < \Omega_{\rm ph}$, the nematic phase is enhanced, while for $\Omega > \Omega_{\rm ph}$ it is suppressed.

To derive these results, we start from a general
model for coupled nematic degrees of freedom and $E_{u}$ phonons.
We are not concerned with the microscopic mechanism of the nematic
transition, and merely describe it by a bosonic field $\phi$ that
transforms as the $B_{1g}$ irreducible representation of the tetragonal
point group. In terms of the electronic field operators $\psi\left(\mathbf{k},\nu \right)$,
the nematic bosonic field is proportional to the quadrupolar charge
density, given by $\phi(\q,\w)\sim\sum_{\k,\nu}(k_{x}^{2}-k_{y}^{2})\psi^{\dagger}(\k+\q/2,\nu+\w/2)\psi(\k-\q/2,\nu-\w/2)$.
We describe the phonons by an optical degenerate mode $\mathbf{X}=(X_1,X_2)$ which couples to an electric field $\boldsymbol{\epsilon} = (\epsilon_1, \epsilon_2)$. For concreteness we assume our system is on a two dimensional lattice, and is described by the coupled action,
\begin{equation}
  \label{eq:action-full}
  \mathcal{\hbar S} = \idx \left[\mathcal{L}_{\rm ph}+\mathcal{L}_{\rm nem}+\mathcal{L}_{\rm int}\right],
\end{equation}
where $\mathcal{C}$ is the Keldysh contour, and
\begin{align}
  \label{eq:nem-action}
  \mathcal{L}_{\rm nem} &= \phi \frac{1}{2\chi_0}(-\pd_t^2 - r ) \phi - \frac{u}{4\chi_0}\phi^4, \\
  \mathcal{L}_{\rm ph} &= \frac{M}{2}\mathbf{X} \left(-\pd_t^2 - \W_{\mathrm{ph}}^2\right) \cdot\mathbf{X} +q \mathbf{X}\cdot\bm{\epsilon} \label{eq:ph-action}
                     ,\\
  \mathcal{L}_{\rm int} &= -\frac{\lambda}{a^2} \phi(x)\left(X_1^2-X_2^2\right)\,.
\label{eq:int-action}
\end{align}
Here, $r$ and $u$ have units of frequency squared and describe a Landau theory for the nematic critical point that occurs at $r=0$. Thus, $\sqrt{r}$
should
be understood as the frequency of the nematic collective mode, which softens at the nematic transition. $M$, $\Wp$, and $q$ describe the mass, resonance frequency, and ionic charge of the $E_u$ mode.
$\lambda$ is the effective coupling and $a = \sqrt{\hbar/ M\Wp}$ is the phonon oscillator length. 
We choose $\chi_0$ such that $\phi$ is dimensionless. In the usual manner, we split $\mathcal{C}$ into backwards ($<$) and forwards ($>$) segments, and perform the Keldysh rotation for the fields, i.e.
$\phi^{\rm cl,q} = (\phi_>\pm \phi_<)/2$.
Then we integrate out the phonon modes and obtain an effective action for the nematic mode,
\begin{equation}
  \label{eq:nem-eff-exact}
  S_{\rm nem} = \sum_{\x_i} \Tr \left[ \frac{1}{2}\Phi \chi_{\rm nem}^{-1}\Phi - \frac{q^2}{2}\mathcal{E} \tau^1 D \tau^1\mathcal{E} \right]
\end{equation}
where the trace is over (Euclidean) time and
\begin{equation}
  \label{eq:D-full-def}
  D = \left[D_0^{-1} - 2\frac{\lambda}{a^2}\tilde\Phi \sigma^z\right]^{-1}-D_{0}.
\end{equation}
Here, $\Phi = (\phi^{\rm cl},\phi^{\rm q})$, $\tilde\Phi = \tau^1\phi^{\rm cl}+\tau^0\phi^{\rm q}$, $\mathcal{E} = (\boldsymbol{\epsilon}^{\rm cl},\boldsymbol{\epsilon}^{\rm q})$, $\boldsymbol{\tau}$ and $\boldsymbol{\sigma}$ are Pauli matrices in Keldysh and coordinate space, respectively, and $D_0$ and $\chi_{\rm nem}$ are the equilibrium propagators of the phonon and nematic modes, whose retarded components are given by the $({\rm cl,q})$--component of the matrix propagators:
\begin{align}
  \label{eq:D0-r}
  D^{R}(\W) \equiv D^{\rm ( cl,q )}_0(\W) &= \frac{M^{-1}}{\W^2 - \Wp^2 +  i \Gamma_{\rm ph} \W} \sigma^0,\\
  \chi^{R}(\w) &= \chi_0(\w^2-r+ i \Gamma_{\rm nem}\w)^{-1}. \label{eq:D0-r2}
\end{align}
$\Gamma_{\rm nem}$ and $\Gamma_{\rm ph}$ are phenomenological damping terms, arising from e.g. Landau damping.

Eq. \eqref{eq:nem-eff-exact} is our first main result. It demonstrates that the phonons modify the nematic action in proportion to the laser intensity $|\mathcal{E}|^2$. To proceed further analytically we expand $D$ perturbatively to second order in $\phi$ (the expansion is also justified when $\lambda/\Wp \ll 1$ and frequencies are not too close to the phonon resonance). The leading contributions to the action are:
\begin{align}
  \label{eq:chi-conts-1}
  \delta S^{(1)} &= -  \frac{\lambda q^2}{a^2}\Tr \mathcal{E} \tau^1 D_0 \tilde{\Phi} D_0 \sigma^z \tau^1 \mathcal{E}, \\
  \delta S^{(2)} &= - \frac{2\lambda^2 q^2 }{a^4}\Tr \mathcal{E}\tau^1 D_0 \tilde{\Phi}D_0 \tilde{\Phi} D_0\tau^1\mathcal{E}  \label{eq:chi-conts-2}.
\end{align}
Eq. \eqref{eq:chi-conts-1} implies that the linear term in $\Phi$ vanishes when the laser excites both $E_u$ modes equally (e.g. for circularly polarized or $45^\circ$ polarized light). Then the only contribution is the quadratic term, Eq. \eqref{eq:chi-conts-2}, which corresponds to a shift $\delta \chi^{-1}$ to the inverse nematic susceptibility $\chi^{-1}$ in Eq. \eqref{eq:D0-r2}.

While the results above are general, we now focus on
the physically interesting case of a nematic mode that is much slower than the phonons $r \ll \Wp^2$, as expected near a nematic phase transition. We also specialize to a nearly monochromatic laser beam with $\epsilon_1 = \epsilon_2=\varepsilon(t)=\varepsilon_0 \cos (\W t) / \sqrt{2}$ and assume a classical laser field $\boldsymbol{\epsilon}^q \approx 0$. The poles of the renormalized retarded nematic susceptibility $ (\tilde{\chi}^{-1})^{R} =  (\chi^{-1})^{R} + (\delta \chi^{-1})^R$ are then given by
\begin{equation}
  \label{eq:chi-poles}
  \w^2 = r - i \Gamma_{\rm nem} \w + \frac{\lambda^2\chi_0}{\hbar\Wp}n_{\rm ph}\left[d^R(\w+\W) + d^R(\w-\W)\right]\,,
\end{equation}
where $n_{\rm ph} = \langle X_1^2 + X_2^2 \rangle/a^2 \approx |D^R(\W)|^2q^2 \varepsilon_0^2 /a^2$, is the average phonon occupation number, and $d^R=M\Wp^2 D^R$ is a  normalized propagator. For details, see the Supplementary Material (SM)~\cite{supplementary}. Eq.~\eqref{eq:chi-poles} has six solutions. To gain insight into them, we consider the classical equations of motion for the coupled phonon-nematic system:
\begin{align}
  \label{eq:classical-eq-1}
  & \ddot{X_i} + \Gamma_{\rm ph} \dot X_i + \left(\Wp^2  \pm \frac{2\lambda}{M a^2} \phi \right) X_i = \frac {q\varepsilon_0} {\sqrt{2}M}\cos\W t\\
  & \ddot{\phi} + \Gamma_{\rm nem}\dot \phi + r\phi +u\phi^3 +\frac{\lambda\chi_0}{a^2}(X_1^2 - X_2^2) = 0.  \label{eq:classical-eq-2}
\end{align}
Substituting the ansatz $\phi(t) = \text{Re}\phi_0 e^{-i\w t}$, the values of $\omega$ that solve the classical equations coincide with the poles of Eq.~\eqref{eq:chi-poles}. Eq.~\eqref{eq:classical-eq-2} is very similar to the three-wave mixing problem, typical of cavity quantum-electrodynamic applications~\cite{Devoret2012}. What is special in our case is that the three-wave mixing term is proportional to $X_1^2 - X_2^2$, i.e. it vanishes unless the nonlinear feedback from $\phi$ is accounted for.
The first set of poles corresponds to a correction of the ``bare'' frequency $\w = \sqrt{r}$ of the nematic mode. If $\sqrt{r} \ll \{ \Wp$, $\Gamma_{\rm ph}$, $\W \}$ is the smallest scale in the problem, we may expand the propagators in Eq.~\eqref{eq:chi-poles} in $\w$ and find
\begin{equation}
  \label{eq:r-shift-1}
  -\w^2 - i (\Gamma_{\rm nem}+\Gamma_{\varepsilon}) \w + (r + r_{\varepsilon}) \approx 0,
\end{equation}
where
\begin{align}
  \label{eq:r_E-def}
  r_{\varepsilon} &= - n_{\rm ph}\frac{2\lambda^2\chi_0}{\hbar\Wp}|d^R(\W)|^2(1-\W^2/\Wp^2),  \\
  \Gamma_{\varepsilon} &= \Gamma_{\rm ph}n_{\rm ph}\frac{2\lambda^2\chi_0}{\hbar\Wp^3}|d^R(\W)|^4\frac{2(\Wp^4-\W^4)-(\Wp^2-\W^2)^2}{\Wp^4}. \label{eq:Gamma_E-def}
\end{align}
This is our second main result, showing that $r$ is shifted either to or away from the nematic transition depending on the detuning between the laser frequency $\Omega$ and the phonon frequency $\Wp$, and in proportion to the pulse intensity $\varepsilon_0^2$.
\begin{figure}
\includegraphics[width=0.95\columnwidth]{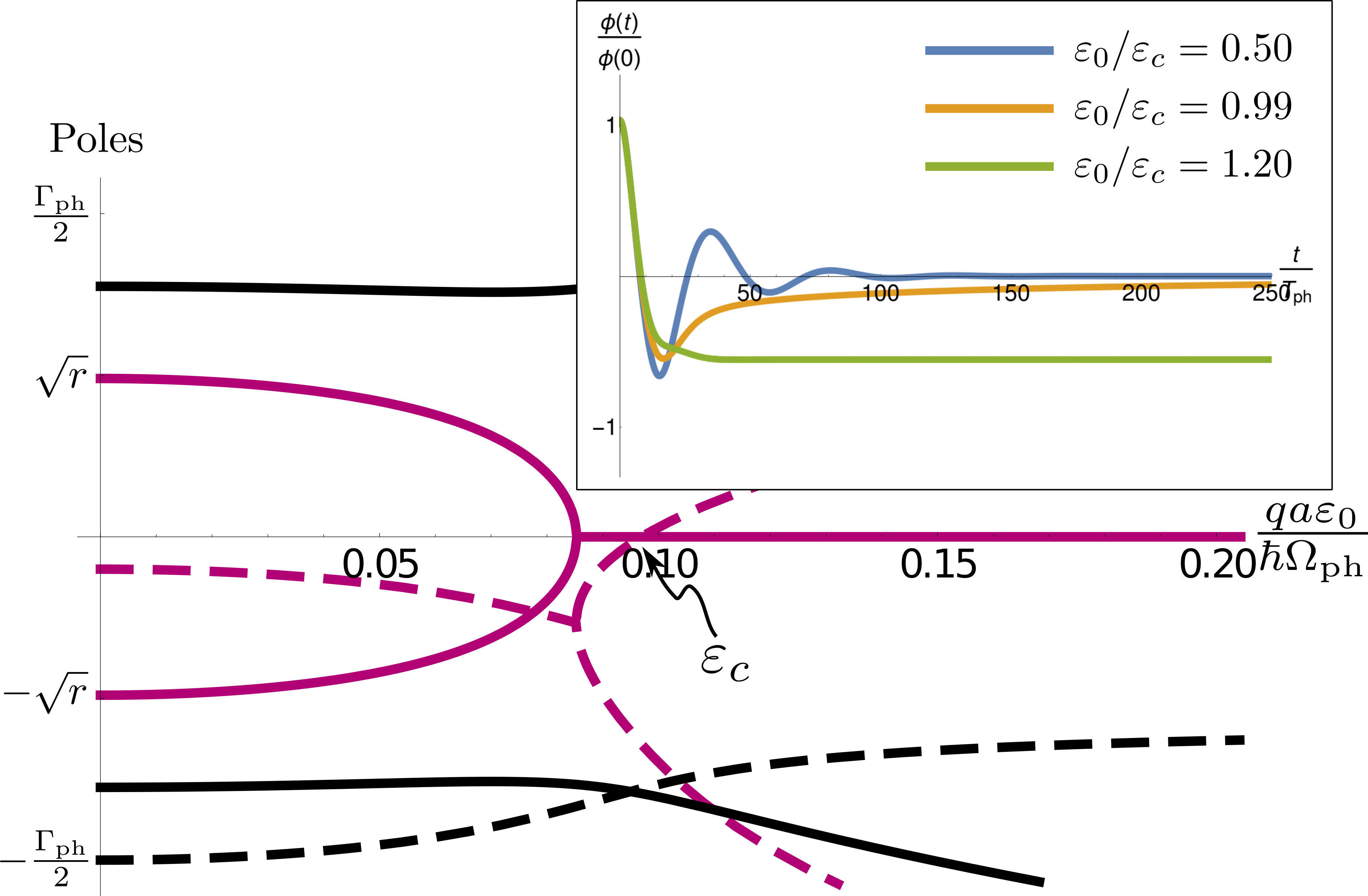}
\caption{Evolution of the low-lying poles of the nematic susceptibility as function of the intensity of the applied electric field $\varepsilon_0$. Solid (dashed) lines denote the real (imaginary) parts of the poles. The magenta trace is a pole that is softening due to the applied field, triggering an instability when its imaginary part
  becomes positive at the critical field strength $\varepsilon_c$.
  The inset depicts the classical evolution of $\phi(t)$ for several $\varepsilon_0$ values. The parameters used are given in the SM~\cite{supplementary}.\label{fig:time-M}}
\end{figure}
Fig.~\ref{fig:time-M} depicts the evolution of these poles for a slightly red-detuned pulse. As $\varepsilon_0$ increases, the nematic mode frequencies soften towards zero, and eventually split and become imaginary. At a critical value $\varepsilon_0=\varepsilon_{c}$, one of these imaginary poles crosses the real axis, signaling an instability towards nematic order, as shown by the time-evolution of the classical solution $\phi(t) = \phi_0 e^{-i\w t}$ depicted in the inset. An equivalent picture holds for a blue-detuned pulse, except that now the system is pushed out of the nematic state (see SM~\cite{supplementary}). 

The figure also displays other poles of $\tilde{\chi}^R$ from Eq.~\eqref{eq:chi-poles}. In addition to the pair of poles discussed above, there are four additional poles corresponding approximately to the poles of the phonon propagators $D^R(\W+\w)$ and $D^R(\w-\W)$, which are therefore purely dynamic and nonperturbative in nature. For
small detuning $\delta = \Wp - \W \ll \W$, two of the poles are of order $\pm 2\Wp$ and do not contribute to the slow dynamics of the nematic mode. The other poles obey
$\w \approx -i\frac{\Gamma_{\rm ph}}{2}\pm
  \delta\sqrt{1-n_{\rm ph}\frac{\lambda^2\chi_0}{\hbar\Wp r}|d_r(\W)|^2}$. Note that the residue of these four additional poles vanish as $\varepsilon_0 \rightarrow 0$.
  
\begin{figure}
\includegraphics[width=0.95\columnwidth]{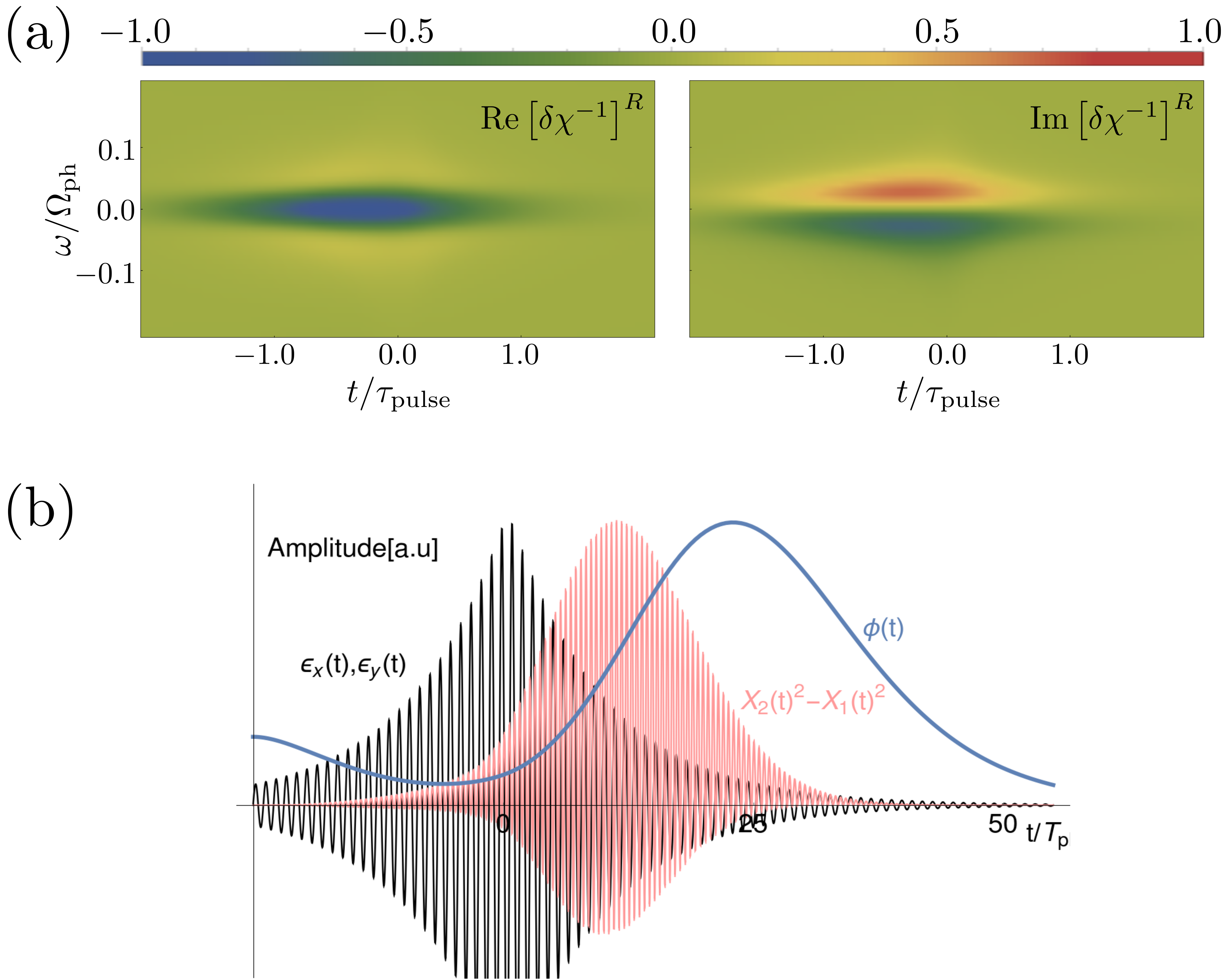}
\caption{Evolution of the nematic order parameter in the case of a laser pulse of duration $\tau_{\mathrm{pulse}}$. (a) Real and imaginary parts of the correction to the nematic susceptibility. (b) Classical evolution of $\phi(t)$ and $X^2_2(t) - X_1^2(t)$ in response to the same pulse. Note that the phonon modes are excited first by the pulse, and only later the nematic field responds. Here, $T_{\rm ph}=2\pi/\Omega_{\rm ph}$. See SM for details and parameters.
}
\label{fig:eff-action}
\end{figure}
The shifting of the nematic mode $r_{\varepsilon}$ in Eq.~\eqref{eq:r_E-def} has a simple physical interpretation. The situation when $\sqrt{r}$ is the smallest frequency in the system corresponds to a type of inverse Born-Oppenheimer problem, where the electronic dynamics is much slower than the ionic (phononic) dynamics. The rapid phonon oscillations can then be integrated out, leaving a quasistatic contribution to the effective nematic
excitation energy.
As long as the lattice symmetry is not explicitly broken by these oscillations --
i.e. the first-order contribution in Eq. \eqref{eq:chi-conts-1} vanishes -- the net effect is a slowdown or speedup of the nematic fluctuations.

While the results above were derived for a monochromatic beam, in actual experiments the laser is pulsed, i.e. $\varepsilon(t)=\varepsilon_0 \cos (\Omega t) e^{-|t|/\tau_{\rm pulse}}$, where $\tau_{\rm pulse}$ controls the duration of the pulse. To show that the main results remain valid in this case, in Fig.~\ref{fig:eff-action}(a) we present the effective shift of the nematic susceptibility $\delta \chi^{-1}(t,\w)$ caused by such a pulse with a red-shift detuning. The real part has a finite contribution at zero frequency for the duration of the pulse, which is equivalent to the shift $r_{\varepsilon}$ of the nematic collective mode in Eq. (\ref{eq:r_E-def}). Similarly, while the pulse is on, the imaginary part has an odd-in-frequency contribution whose coefficient is equivalent to $\Gamma_{\varepsilon}$ of the nematic collective mode in Eq. (\ref{eq:Gamma_E-def}).

These results support a feasible protocol for a quantum quench near
a putative nematic quantum critical point (QCP), consisting of a pulse with frequency $\W \sim \Wp$ and length $\tau_{\rm pulse}$, which excites both components of the $E_u$ mode equally. 
The other relevant time scales for this protocol are the characteristic nematic time $\W_{\text{nem}}^{-1} \sim \text{min}(r^{-1/2},\Gamma_{\rm nem} r^{-1})$, the phonon decay time $\Gamma_{{\rm ph}}^{-1}\gg\Omega_{{\rm ph}}^{-1}$,
and $\tau_{\mathrm{el}}$, the electronic heating time. To establish
coherent dynamics, the latter must be the longest time scale of the
problem, i.e. $\tau_{\mathrm{el}}\gg\Omega_{\mathrm{nem}}^{-1},\,\Gamma_{{\rm ph}}^{-1}$.
We expect this to be the case because, in contrast to traditional
pump-probe experiments,
our method excites the lattice directly while electronic heating occurs indirectly~\cite{Allen1987}.
The excitation of the phonon mode occurs on a timescale $\tau_{\rm ph} \sim \Gamma_{\rm ph}^{-1}$ for a resonant laser pulse or $\tau_{\rm ph}\sim \Wp^{-1}$ for an off-resonant pulse. If 
$\tau_{\rm ph} \gg \W_{\rm nem}^{-1}$, as is expected for an on-resonance excitation fairly far from the nematic transition, the modification of the nematic behavior is quasistatic and the pole of the nematic susceptibility is dissipative.
For $\tau_{\rm ph} \ll \W_{\text{nem}}^{-1}$, as expected near the critical point or for a sufficiently off-resonant laser pulse, the induced softening of the nematic mode [Eq.~\eqref{eq:r_E-def}] is essentially instantaneous. This results in a quantum quench of the nematic instability.
Fig.~\ref{fig:eff-action}(b) shows the classical evolution of both the nematic and phononic degrees of freedom in response to such a pulse. In this particular simulation, $\phi(t)$ (blue curve) starts non-zero and decays until the pulse (black curve) kicks in and enhances the nematic order parameter well above its initial value, but only after the phonon mode has been excited (red curve). 

To show that these conditions can be realized in actual systems, we
consider the case of FeSe~\cite{Coldea2017,Boehmer2017}. This iron-based
superconductor ($T_{\mathrm{SC}}\approx8$ K) displays an electronic
nematic transition at $T_{{\rm {nem}}}\approx90$ K, that is suppressed
to zero upon S doping, suggesting a putative metallic nematic QCP~\cite{Reiss2017,Shibauchi2016}. Because FeSe$_{1-x}$S$_{x}$ does
not display long-range magnetic order at ambient pressure, it is an
ideal system for studying the interplay of nematicity and superconductivity.
We estimate~\cite{supplementary} from existing data
on FeSe and related compounds~\cite{Nelson2002defects, Zakeri2017, Fedorov2016, Subedi2008, Wang2011, Wang2012, Nakajima2017, Phelan2009,Ksenofontov2010,  Homes2016, Gnezdilov2013, Massat2016, Blumberg2017, Hu2016, Luo2012, Luo2012a, Patz2014} that $\hbar\W_{{\rm ph}}\sim30{\rm {meV}}$,
$\hbar\Gamma_{{\rm ph}}\sim0.5-1.5{\rm {meV}}$, and $\hbar\W_{{\rm {nem}}}\sim10{\rm {meV}}$
away from the transition, softening further as one approaches it.
In addition, we estimate $\tau_{{\rm el}}^{-1}<0.25{\rm {meV}}$ away
from the transition, which can get as small as $0.04{\rm {meV}}$
approaching the transition. These estimates show that the quantum
quench regime is achievable in FeSe.

We use these parameters to also estimate the coupling $\lambda$ and,
thus, the expected maximum shift in the nematic susceptibility. As
shown in
Eq. \eqref{eq:classical-eq-1},
static nematic order $\langle\phi\rangle\neq0$
splits the frequency of the $E_{u}$ mode by $\hbar\Delta\Omega\approx 2\lambda\langle\phi\rangle$.
We estimate
$\langle\phi\rangle$
from 
the elliptical distortion of the Fermi surface measured
by ARPES, $\langle\phi\rangle\approx\Delta k_{F}/k_{F}$. This leads
to $\lambda/\hbar\W_{{\rm ph}}\sim0.04$.
The maximum shift of the
nematic transition temperature can then be obtained by a Lindemann
criterion argument: the maximum possible occupation $n_{{\rm ph}}^L$
will have $\langle X^{2}\rangle\approx a^{2}n_{{\rm ph}}^L\approx c_{L}^{2}\ell^{2}$,
where $\ell=3.7\AA$ is the FeSe lattice constant, and $c_L$ is usually a fraction of the order $c_L=0.15$
~\cite{Nelson2002defects}. Using these values and a conservative estimate
$n_{\rm ph} = 0.15 n_{\rm ph}^L$ we find 
a shift of
$ 90 $K
in the nematic transition temperature
(about $4$K per phonon),
indicating the experimental feasibility of a quantum quench
both to study and to control the nematic phase over a wide temperature range. 
Since $T_S = 90K$ in undoped FeSe, this range is enough to move pristine samples to the vicinity of the QCP at low temperatures.

In conclusion, we showed that the nonequilibrium excitation of the
infrared-active $E_{u}$ phonon mode present in tetragonal systems
mediates electronic interactions in the nematic channel.
Besides establishing a robust protocol for light control of nematicity,
this result unveils a promising and experimentally feasible avenue
to induce a quantum quench across the transition of correlated materials
that display nematic order. Because nematic fluctuations are intimately
connected to other electronic instabilities, most notably superconductivity
and magnetism, the nonequilibrium excitation of nematic fluctuations
may also be used to drive transient states with different types of
electronic orders. In particular, suppressing nematicity can indirectly enhance superconductivity, since these are known competing orders \cite{Nandi2010}.
\begin{acknowledgments}
We are grateful to M. Sentef, A. Chubukov,
A. Kamenev, P. Orth, M. Gastiasoro,
and M. Sch{\"u}tt
for fruitful discussions,
and to C. Giannetti and Y. Albeck for instruction on ultrafast experimental techniques.
This work (R.M.F. and M.H.C.) was supported
by the by U.S. Department of Energy, Office of Science, Basic Energy Sciences, under Award DE-SC0012336.
\end{acknowledgments}


\end{document}


\title{Supplementary material for ``Laser-induced
  control of an electronic nematic quantum phase
transition''}

\author{Avraham Klein}

\author{Morten H. Christensen}

\author{Rafael M. Fernandes}

\affiliation{School of Physics and Astronomy, University of Minnesota, Minneapolis. MN
55455}

\maketitle
\providecommand{\ph}{\mbox{ph}} \providecommand{\x}{\bv{x}} \providecommand{\X}{\bv{X}}
 \newcommand{\angs}{\mbox{\AA}}

 
\section{Effective action for the nematic mode}
\label{sec:deta-regard-induc}

In this section we provide details of the derivations leading to the results in the main text. We provide the general expression for the frequency dependent effective action induced by an arbitrary pulse, including all Keldysh components. We also provide results for a more general set of parameters than those presented in the main text. In particular, we provide further details concerning the melting of nematic order using a blue-detuned laser.

\subsection{Detailed derivation of the correction to the nematic effective action}
\label{sec:deta-deriv-corr}

We first derive the first- and second-order corrections to the nematic action, Eqs.~(9) and (10) in the main text. Our starting point is the effective action for the nematic mode, Eq. (5) of the main text. The first-order induced correction is
\begin{eqnarray}
	\delta S^{(1)} = -\frac{\lambda q^2}{a^2}\text{Tr}\left[\mathcal{E} \tau^1 D_0\tilde{\Phi}D_0 \sigma^z \tau^1 \mathcal{E} \right].\label{eq:1st_order_eq}
\end{eqnarray}
We recall that $\boldsymbol{\sigma}$ denote Pauli matrices in coordinate-space while $\boldsymbol{\tau}$ denote Pauli matrices in Keldysh-space, and that the phonon propagator has the form
\begin{eqnarray}
	D_0 = \begin{pmatrix}
		D^K & D^R \\
		D^A & 0
	\end{pmatrix} \otimes \sigma^0,
\end{eqnarray}
where $D^{R(A)}$ is the retarded (advanced) component, and $D^{K}$ is the Keldysh component. The vanishing of $\delta S^{(1)}$ for $\varepsilon_1=\varepsilon_2$ (circularly or $45^{\circ}$ polarized light) is a direct consequence of the trace over $\sigma^z$, which yields a term proportional to $\varepsilon_1 - \varepsilon_2$.

The second-order term is
\begin{eqnarray}
  \label{eq:chi-2-supp-1}
	\delta S^{(2)} = -\frac{2q^2\lambda^2}{a^4}\text{Tr}\left[\mathcal{E}\tau^1 D_0 \tilde{\Phi} D_0 \tilde{\Phi} D_0 \tau^1 \mathcal{E} \right].
\end{eqnarray}
We assume that the electric field is purely classical, $\bm{\epsilon}^q = 0$, and write out Eq.~\eqref{eq:chi-2-supp-1} explicitly:
\begin{align}
  \label{eq:chi-2-supp-time}
  \delta S^{(2)} &= -\frac{2q^2\lambda^4}{a^4}\int\mathrm{d}t_1\cdots\mathrm{d}t_{4}~~\varepsilon^{cl}(t_1)D^A(t_1 - t_2) \nonumber\\
           &\qquad\qquad           \begin{pmatrix}
                        \phi^{cl}(t_2) & \phi^{q}(t_2)
                      \end{pmatrix}\cdot
                                         \begin{pmatrix}
                                           0 & D^A(t_2-t_3) \\
                                           D^R(t_2-t_3) &  D^K(t_2-t_3)
                                         \end{pmatrix}\cdot
                                                          \begin{pmatrix}
                                                            \phi^{cl}(t_3) \\
                                                            \phi^{q}(t_3)
                                                          \end{pmatrix}
  D^R(t_3-t_4) \varepsilon^{cl}(t_4) \\
                    &= -\frac{2q^2\lambda^4}{a^4}\int \frac{d\omega_1 d\omega_2 d\nu}{(2\pi)^3}\varepsilon^{cl}(\omega_1-\nu)D^A(\nu-\omega_1)  D^R(\nu-\omega_2) \varepsilon^{cl}(\nu-\omega_2) \nonumber\\
                    &\qquad\qquad\qquad\qquad
                        \begin{pmatrix}
                        \phi^{cl}(-\omega_1) & \phi^{q}(-\omega_1)
                      \end{pmatrix}\cdot
                                             \begin{pmatrix}
                                           0 & D^A(\nu) \\
                                           D^R(\nu) & D^K(\nu)
                                         \end{pmatrix}\cdot
                                                          \begin{pmatrix}
                                                            \phi^{cl}(\omega_2) \\
                                                            \phi^{q}(\omega_2)
                                                          \end{pmatrix}. \label{eq:chi-2-supp-freq}
\end{align}
Eq. \eqref{eq:chi-2-supp-time} gives a correction to the inverse nematic susceptibility $\delta \chi^{-1}$, which can be computed for any pulse shape. We discuss it in more details later in this section. To obtain Eq.~(11) of the main text we specialize to a monochromatic beam,
\begin{equation}
  \label{eq:mono-field}
  \boldsymbol{\varepsilon}^{cl}(t) = \varepsilon_0\hat{\mathbf{n}} \cos\Omega t
\end{equation}
where $\hat{\mathbf{n}}$ is a unit vector in coordinate space, chosen such that the first-order term vanishes. Plugging the Fourier transform of Eq.~\eqref{eq:mono-field} into Eq.~\eqref{eq:chi-2-supp-freq} we find the particular action term from which we can read out the retarded nematic susceptibility:
\begin{eqnarray}
  \label{eq:mono-action}
	\delta \tilde{S}^{(2)} = -\frac{q^2 \lambda^2 \varepsilon_0^2}{2a^4} |D^R(\Omega)|^2 \int\frac{d\omega}{2\pi}\phi^{q}(-\omega) \left[D^{R}(\omega + \Omega) + D^{R}(\omega - \Omega) \right] \phi^{cl}(\omega)\,.
\end{eqnarray}
Here we have neglected terms with a large frequency transfer, of order $2\W_{\rm ph}$. Neglecting these terms is justified if the relevant frequencies of the nematic system, i.e. $\sqrt{r},\Gamma_{\rm nem}$ are smaller than $\Omega_{\rm ph}$. 
Combining this with the bare form of the nematic susceptibility yields Eq. (11), which we reproduce here:
\begin{eqnarray}
  0 = \omega^2 + i\Gamma_{\text{nem}}\omega - r - \frac{\lambda^2 \chi_0}{\hbar \Omega_{\rm ph}}n_{\rm ph} \left[d^{R}(\omega+ \Omega)+ d^{R}(\omega - \Omega) \right]\,,
\end{eqnarray}
Here, as in the main text, $n_{\rm ph} = |D^R(\Omega)|^2 q^2\varepsilon_0^2 /a^2$, $d^R = M \Omega_{\rm ph}^2 D^R$, and $a^2 = \hbar / M \Omega_{\rm ph}$.

\subsection{Evolution of the nematic propagator's poles for generic parameters}
\label{sec:evol-nemat-prop}

In this section we will consider other representative solutions to Eq.~(11) of the main text that were not discussed in the manuscript. Specifically, we will relax the conditions that $\sqrt{r}$ is the smallest energy scale in the problem.
We solve the equation
\begin{eqnarray}
	\omega^2 = r - i \Gamma_{\rm nem} \omega + \frac{\lambda^2 \chi_0}{\hbar \Omega_{\rm ph}}n_{\rm ph}\left[d^R( \omega + \Omega) + d^R(\omega -\Omega) \right]\label{eq:poles_eq}
\end{eqnarray}
and interpret the solution using the classical equations of motion
\begin{eqnarray}
  \ddot{X}_{i} + \Gamma_{\rm ph} \dot{X}_i + \left(\Omega^2_{\rm ph}
  \pm \frac{2\lambda}{Ma^2}\phi \right) X_i &=& \frac{q}{M}\epsilon_i \cos{\Omega t}\label{eq:phonon_eom} \\
  \ddot{\phi} + \Gamma_{\rm nem} \dot{\phi} + r\phi + u \phi^3
  +
  \frac{\lambda\chi_0}{a^2}\left(X_1^2 - X_2^2 \right) &=& 0\,.\label{eq:nem_eom}
\end{eqnarray}

In the figures that follow we present the numerical solution of the pole equation, Eq.~\eqref{eq:poles_eq} and of the classical equations of motion, Eqs.~\eqref{eq:phonon_eom} and \eqref{eq:nem_eom}, for several representative situations. We first present the solutions for the disordered phase $r > 0$, when the beam pushes the system towards the ordered phase, and then present the solutions for the case $r < 0$, when the beam pushes the system out of the ordered phase. For completeness we also include the case considered in the main text and cite the specific parameter values used to produce the figures in the manuscript.
To reduce the number of parameters in Eqs.~\eqref{eq:poles_eq}--\eqref{eq:nem_eom} we perform the following rescalings: $\lambda \rightarrow \lambda / \hbar$, $\hbar \chi_0 \rightarrow \chi_0$, $X_i \rightarrow X_i /a$, and $q \epsilon_i / Ma \rightarrow \epsilon_i$. In all cases, we measure frequencies in units of $\Omega_{\rm ph} =1$ and set $\lambda = 0.01$, $\chi_0=1$, and $u=1$.

Before continuing, a comment is in order concerning the solution of the classical equation of motion for $\phi$. The classical solution requires a non-zero initial value of the nematic order parameter $\phi(t=0)=\phi(0)$, otherwise there is no time-evolution.
In the figures below we have chosen $\phi(0) = 0.031$.
It is easy to see from the figures, e.g. Fig.~\ref{fig:large_r_small_detuning}, that the eventual steady-state configuration of the system is not determined by this initial condition.
In an actual experiment, $\phi(0)$ represents the instantaneous local value of the nematic field, and not the equilibrium value. A non-zero initial value is a consequence of either fluctuations or an electromagnetic field whose components do not obey $\varepsilon_1 = \varepsilon_2$ exactly [see Eq.~\eqref{eq:1st_order_eq}].

The specific cases shown here consider various limits with either $\sqrt{r} < \Gamma_{\rm ph}$ or $\sqrt{r} \gg \Gamma_{\rm ph}$. They also cover cases where the detuning of the laser, $|\Omega_{\rm ph} - \Omega|$, is either comparable to or much larger than the phonon damping, $\Gamma_{\rm ph}$.
In addition, we show solutions when $\Gamma_{\text{nem}}$ is large, of the order of $\Omega_{\rm ph}$.
\subsubsection{Disordered phase ($r>0$)}
\label{sec:disordered-phase-r0}

\begin{figure}
\includegraphics[width=0.45\textwidth]{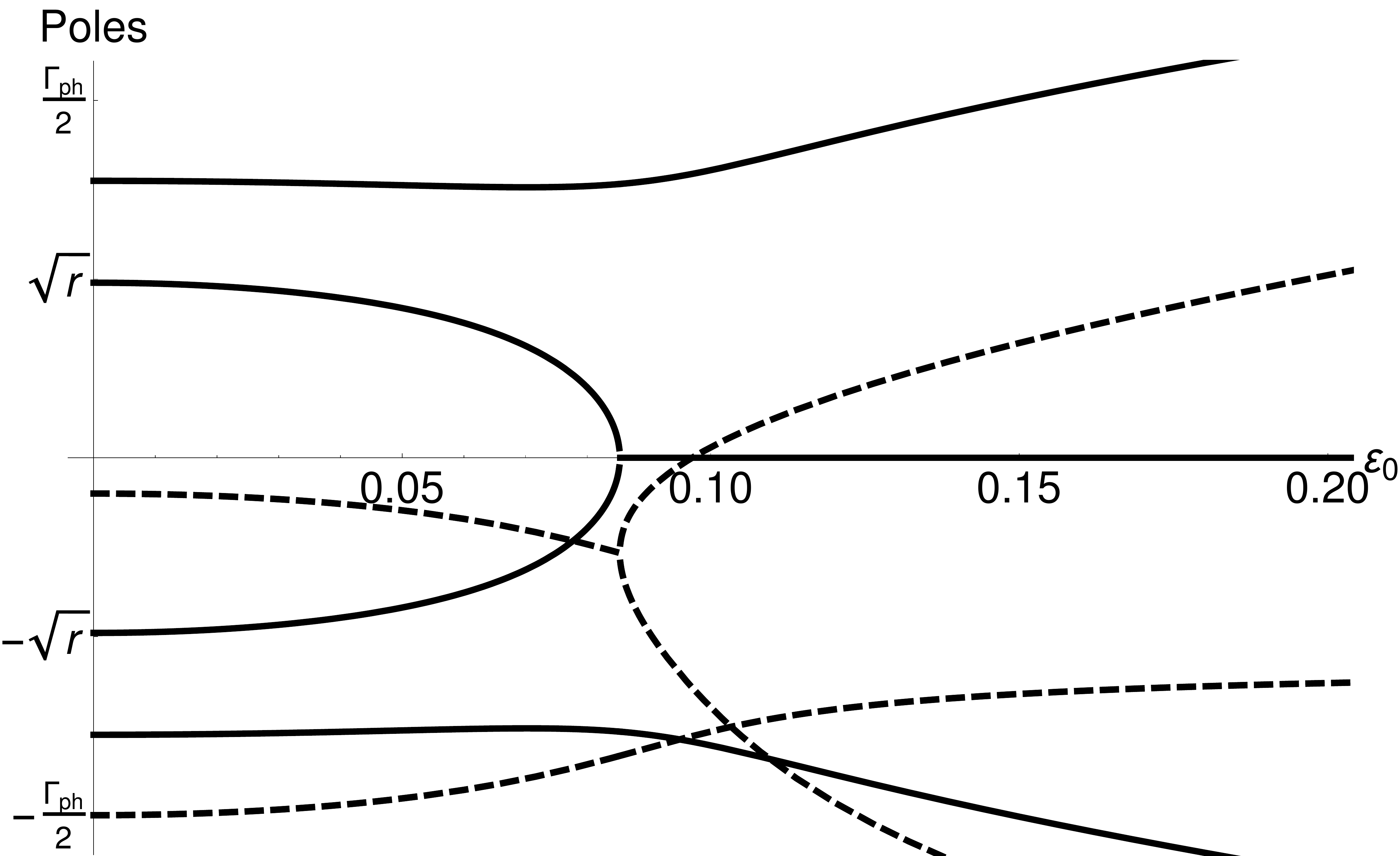}\hspace{2mm}\includegraphics[width=0.45\textwidth]{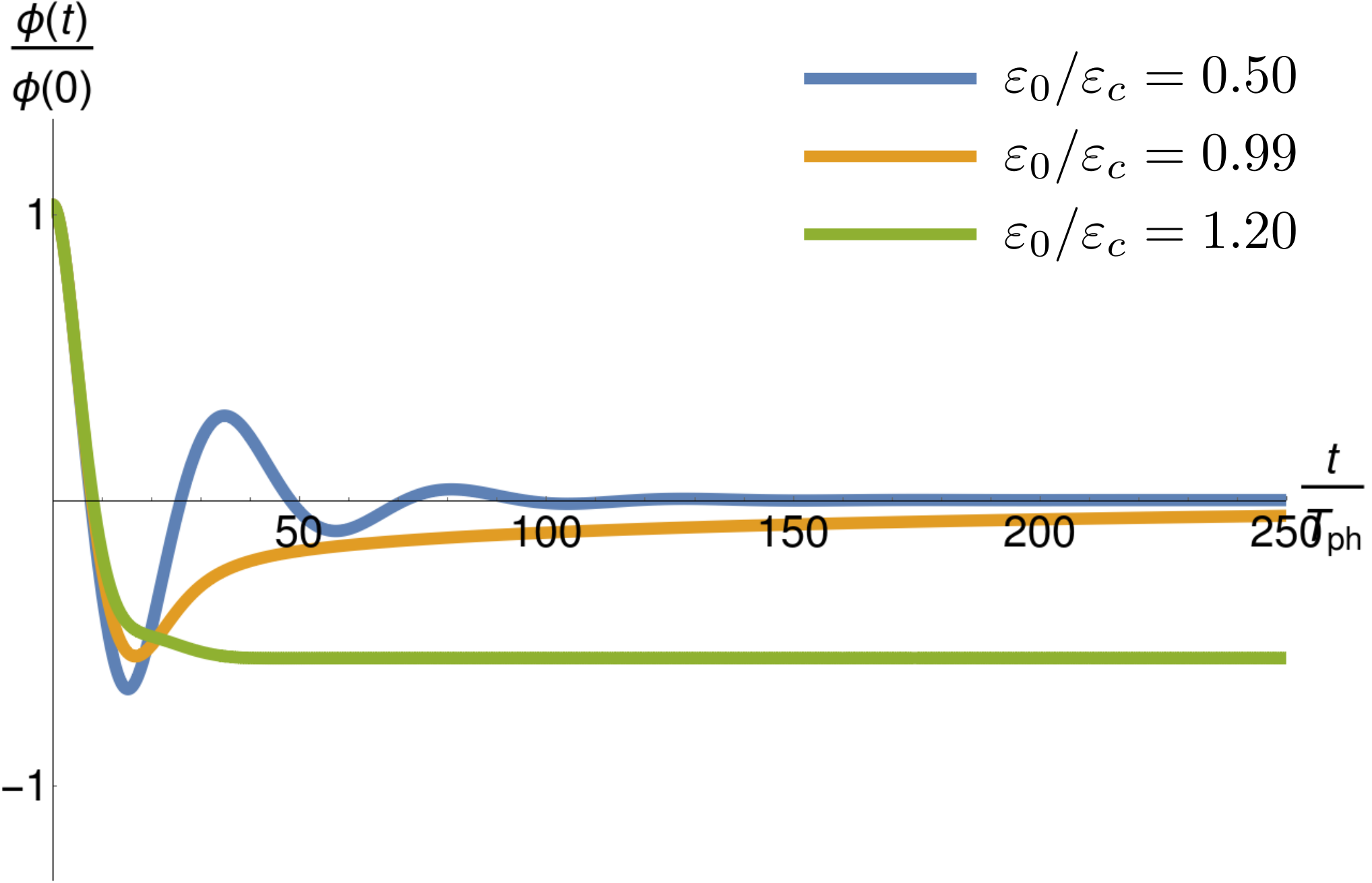}
\caption{\label{fig:small_r_small_detuning}Evolution of the poles of the nematic propagator (left) and classical evolution of the nematic order parameter (right) for the case presented in the main text. Here $\sqrt{r}=0.025$, $\Gamma_{\rm ph} = 0.1$, and $\Gamma_{\rm nem}=0.01$. The laser is red detuned from the phonon resonance and is set at $\Omega = 0.96$ (recall $\Omega_{\rm ph}=1$). The critical field is $\varepsilon_{0,c}\approx 0.097$.}
\end{figure}

\begin{figure}
\includegraphics[width=0.45\textwidth]{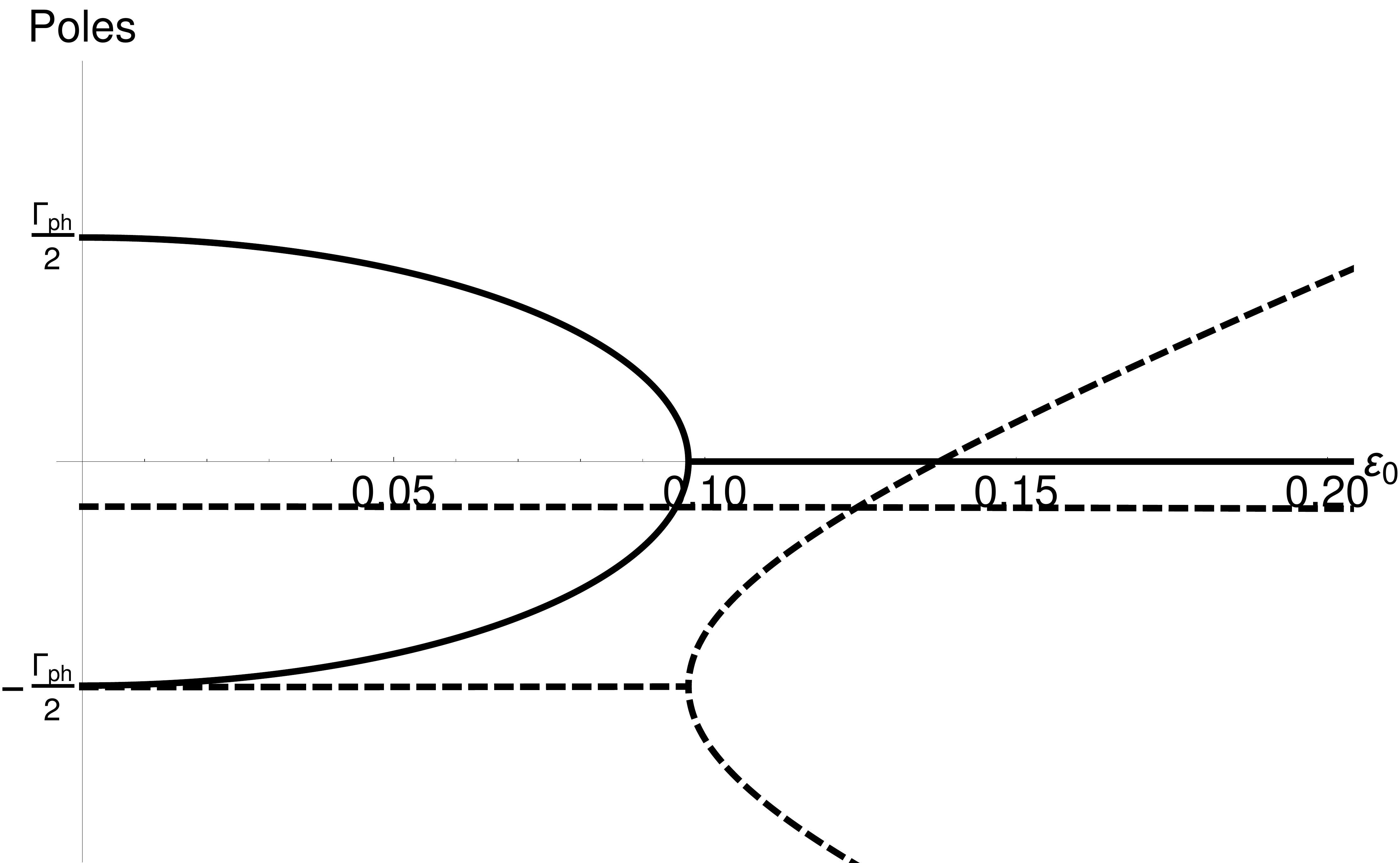}\hspace{2mm}\includegraphics[width=0.45\textwidth]{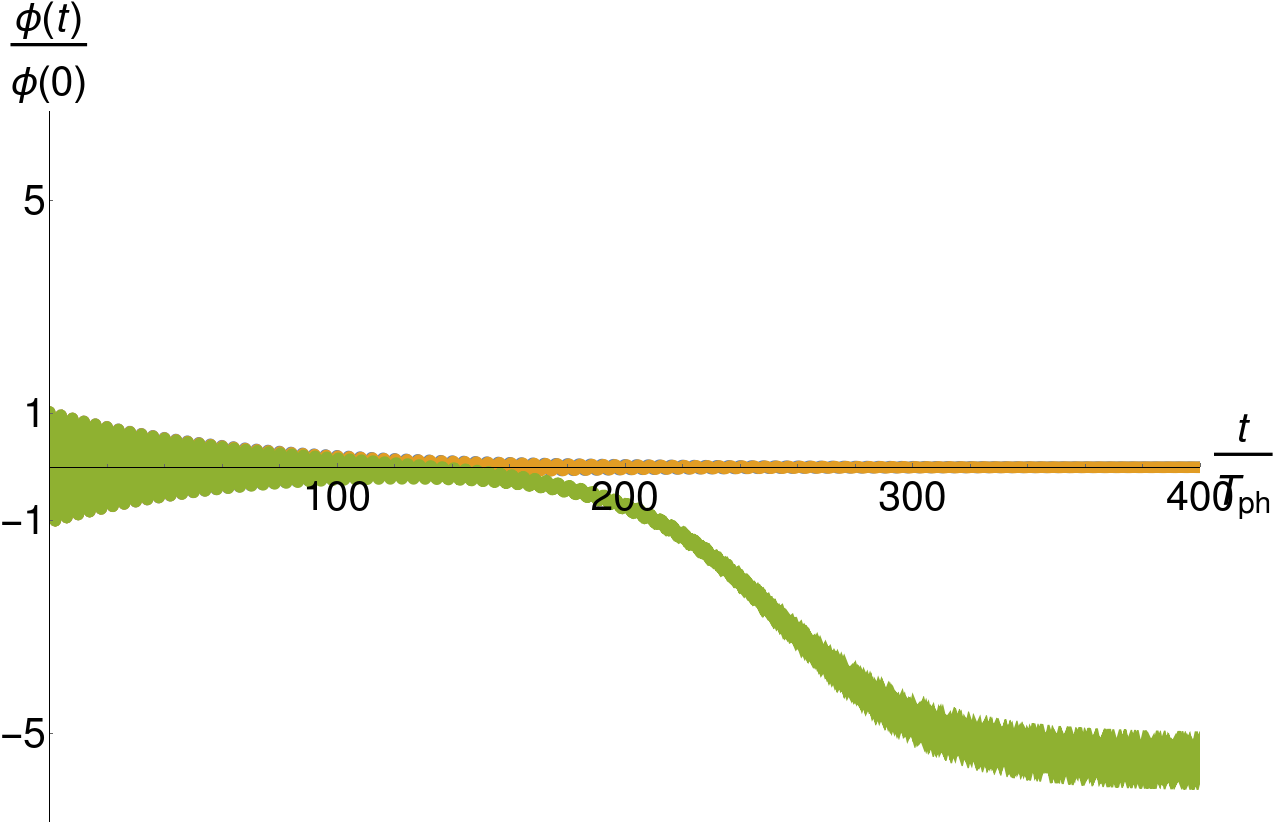}
\caption{\label{fig:large_r_small_detuning}Evolution of the poles of the nematic propagator (left) and classical evolution of the nematic order parameter (right) when $\sqrt{r} \gg \Gamma_{\rm ph}$. Specifically, $\sqrt{r}=0.25$, $\Gamma_{\rm ph}=0.025$, $\Gamma_{\rm nem}=0.005$, and the laser is red-detuned, with frequency $\Omega=0.9875$. The critical field is $\varepsilon_{0,c} \approx 0.138$. The color code in the right panel is the same as in Fig. \ref{fig:small_r_small_detuning}; the blue curve is
  hidden behind
  the yellow one.}
\end{figure}

\begin{figure}
\includegraphics[width=0.45\textwidth]{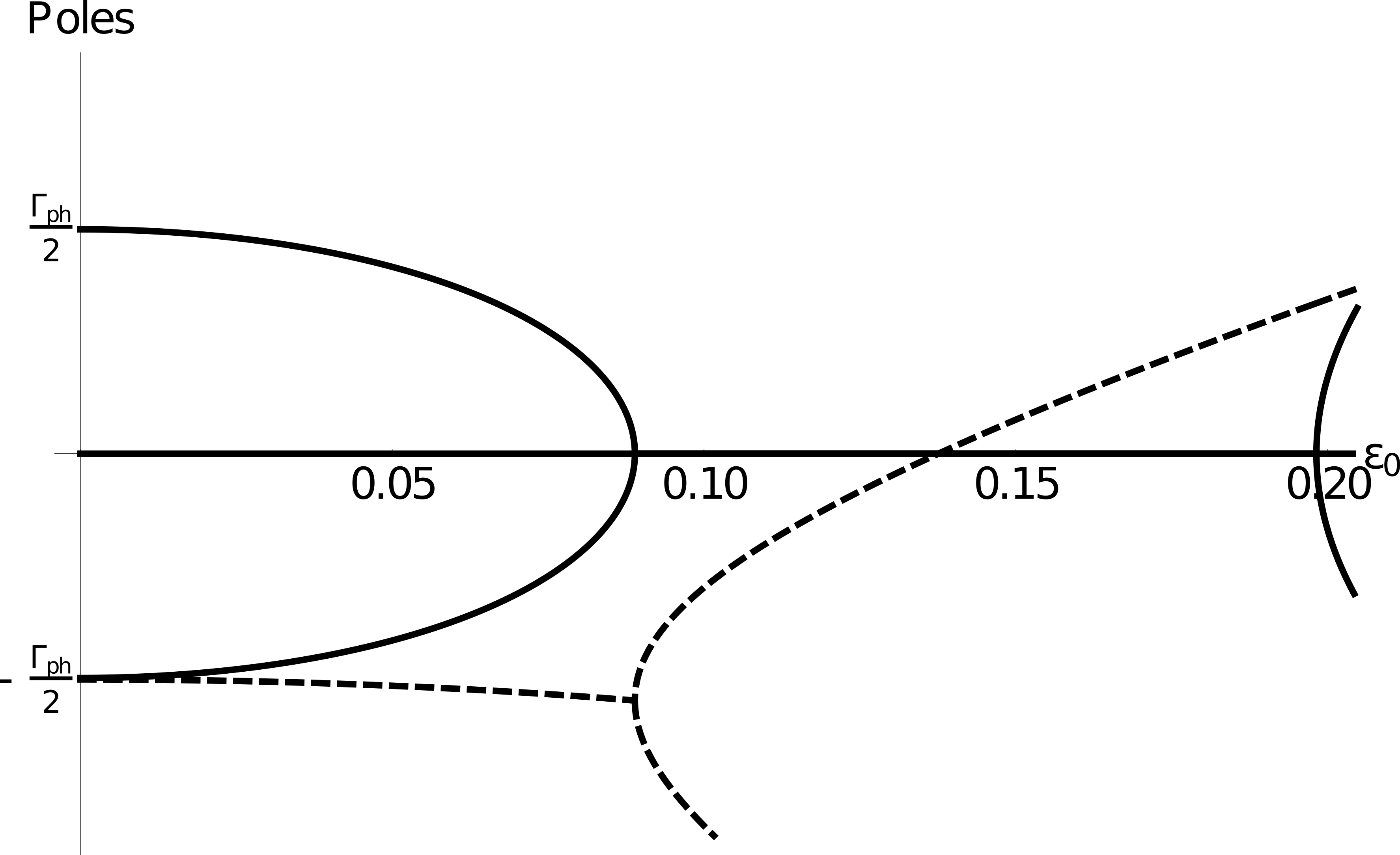}\hspace{2mm}\includegraphics[width=0.45\textwidth]{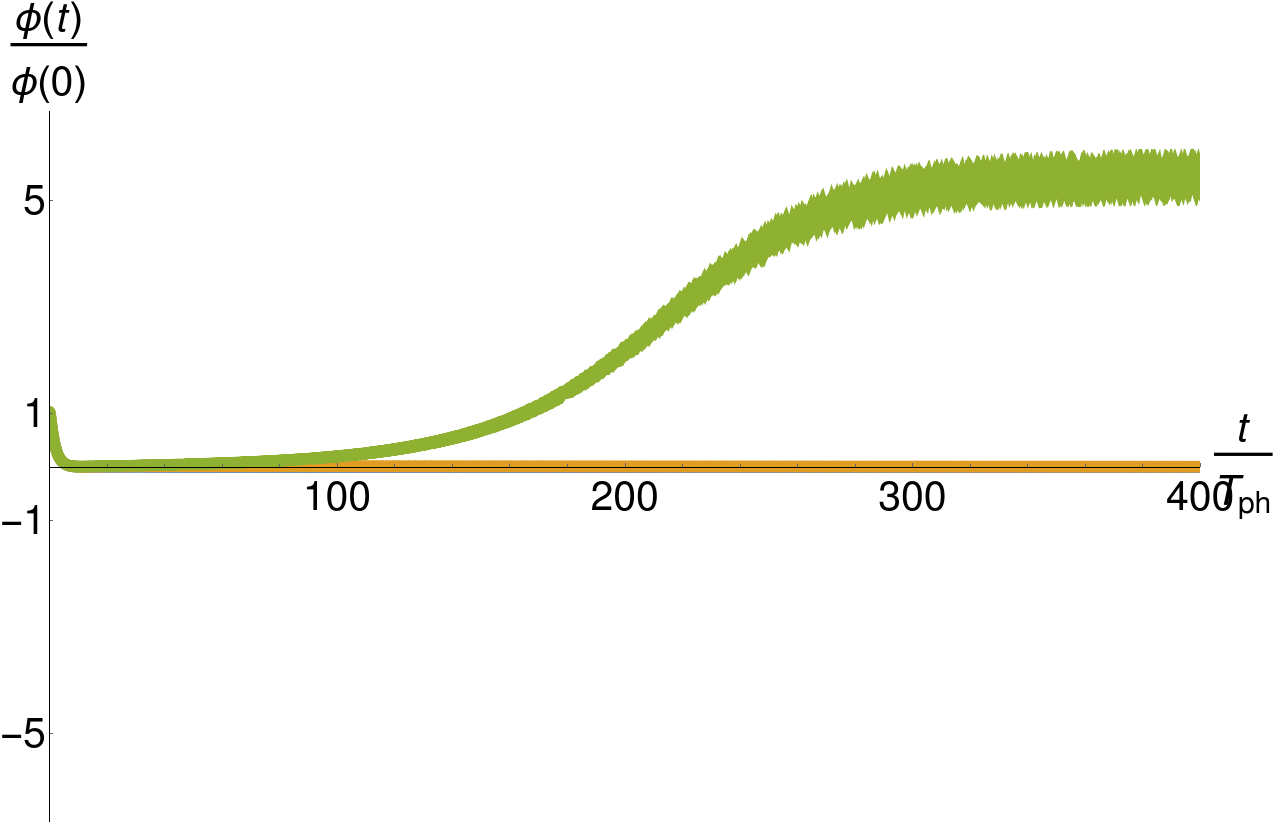}
\caption{\label{fig:large_r_small_detuning_large_damping}Evolution of the poles of the nematic propagator (left) and classical evolution of the nematic order parameter (right) for the case where the Landau damping of the nematic mode is comparable to the phonon resonance, $\Omega_{\rm ph} \sim \Gamma_{\rm nem}$. Specifically, we choose $\sqrt{r}=0.25$, $\Gamma_{\rm ph}=0.025$, $\Gamma_{\rm nem}=0.8$, and $\Omega = 0.9875$. The critical field is $\varepsilon_{0,c}\approx 0.138$ (identical to the case shown in Fig.~\ref{fig:large_r_small_detuning}). The color code in the right panel is the same as in Fig. \ref{fig:small_r_small_detuning}; the blue curve is
  hidden behind
  the yellow one.}
\end{figure}

\begin{figure}
\includegraphics[width=0.45\textwidth]{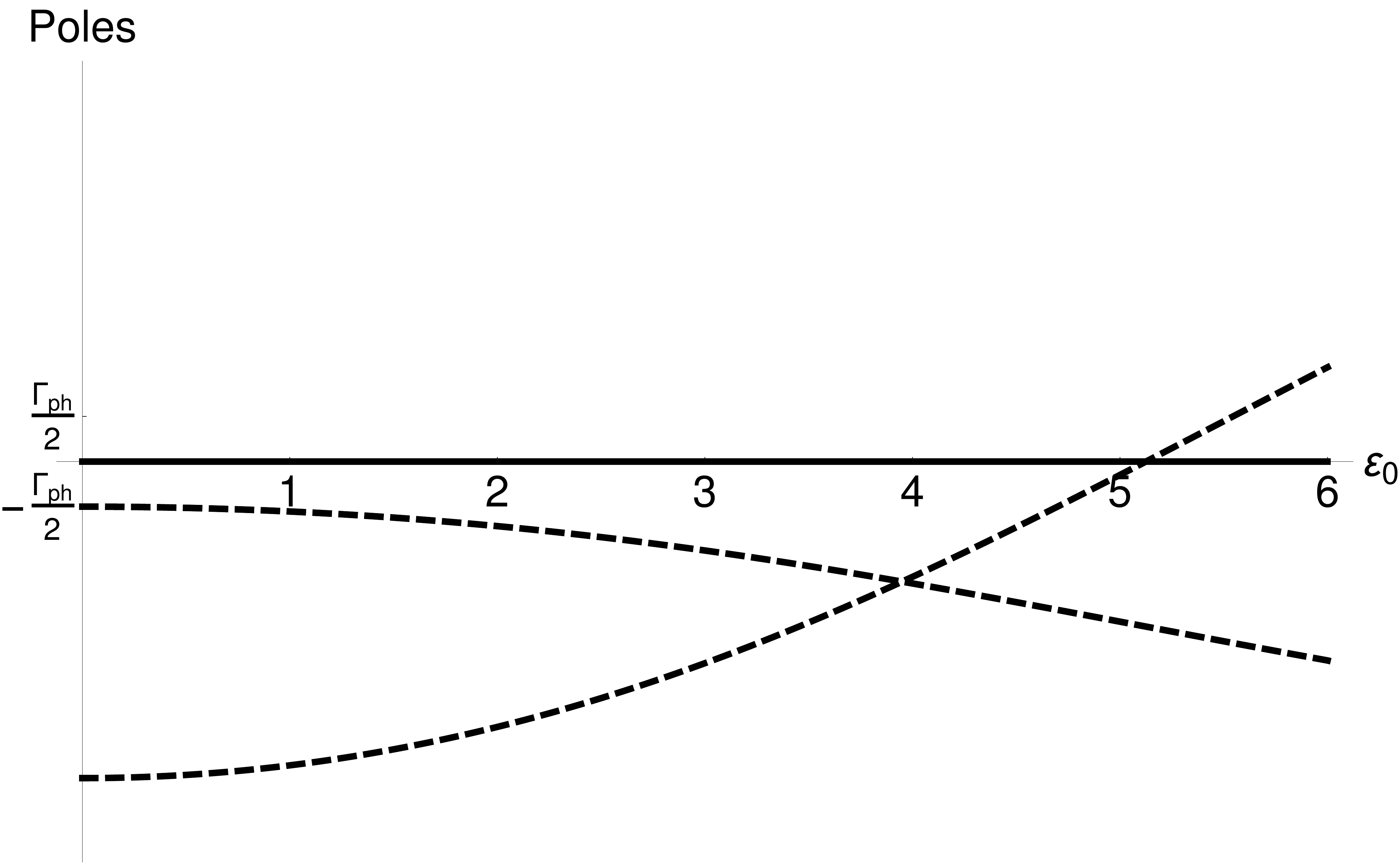}\hspace{2mm}\includegraphics[width=0.45\textwidth]{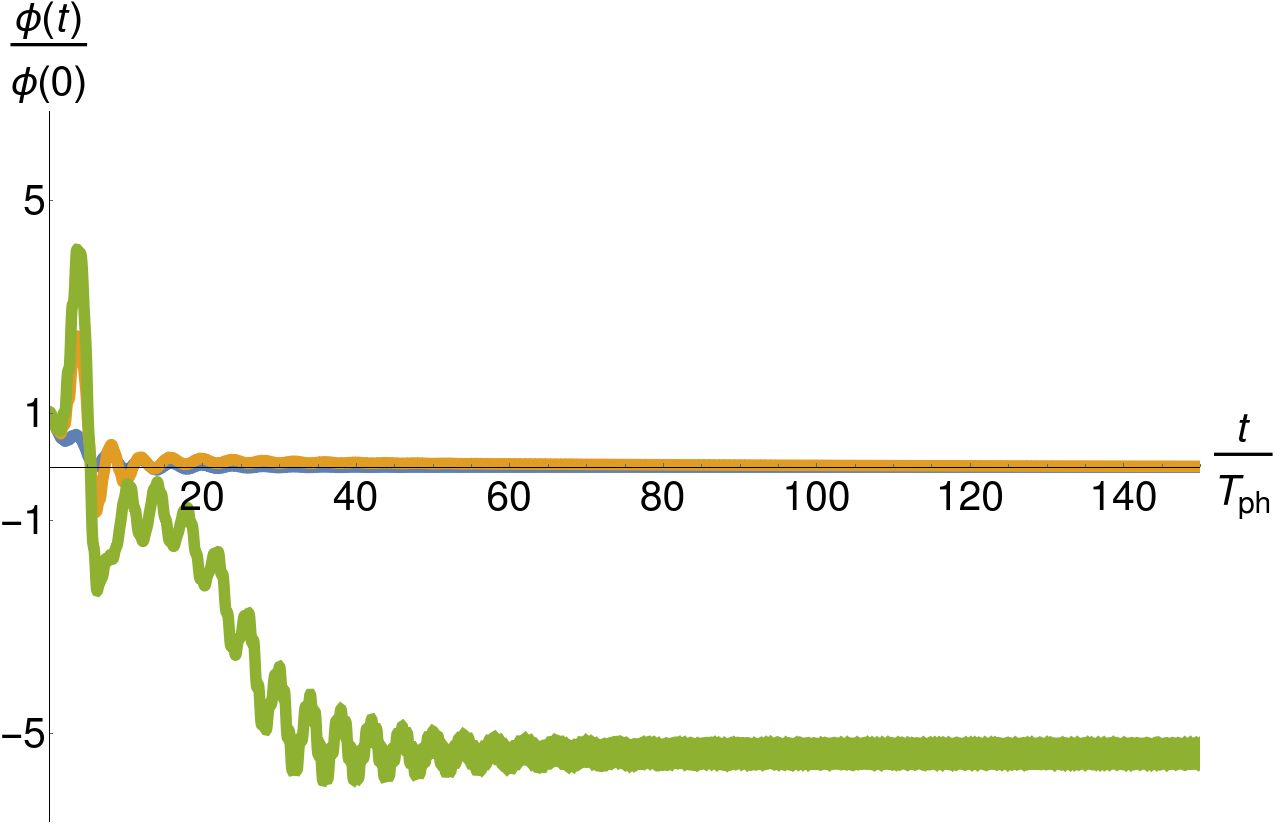}
\caption{\label{fig:large_r_large_detuning_large_damping}Evolution of the poles of the nematic propagator (left) and classical evolution of the nematic order parameter (right) for the case where the Landau damping of the nematic mode is comparable to the phonon resonance and the laser detuning is large compared to the phonon damping, $\Omega_{\rm ph} - \Omega \gg \Gamma_{\rm ph}$. The values used here are $\sqrt{r}=0.25$, $\Gamma_{\rm ph}=0.025$, $\Gamma_{\rm nem}=0.8$, and $\Omega = 0.75$. The critical field is $\varepsilon_{0,c}\approx 5.13$. The color code in the right panel is the same as in Fig. \ref{fig:small_r_small_detuning}.}
\end{figure}

Figs.~\ref{fig:small_r_small_detuning}--\ref{fig:large_r_large_detuning_large_damping} are for a red-detuned beam ($\Omega<\Omega_{\mathrm{ph}}$), implying that the system is driven towards the nematic phase. Specific parameters for each case are shown in the figure captions; solid (dashed) lines denote the real (imaginary) parts of the poles. Note that, in all cases, there is steady-state nematic order for large enough electric fields, when the imaginary part of one of the poles become positive.

\subsubsection{Ordered phase ($r < 0$)}
\label{sec:ordered-phase-r}

In the ordered phase, the steady-state value of the nematic mode at zero field is nonzero, making the solutions of Eqs. \eqref{eq:phonon_eom} and \eqref{eq:nem_eom} more complex. However, as long as $\phi$ is small, we can still linearize the equations and obtain analytic solutions. For concreteness, consider the case where $|r|$ is the smallest scale in the problem, i.e. we are approaching the nematic transition from the ordered side. Then the equation of motion for $\phi$ becomes
\begin{equation}
  \label{eq:nem-eom-ordered}
   \ddot{\phi} + (\Gamma_{\rm nem} + \Gamma_{\varepsilon}) \dot{\phi}  +(r + r_\varepsilon)\phi + u \phi^3 = 0\,,
\end{equation}
where $\Gamma_\varepsilon,r_\varepsilon$ were defined in Eqs. (15)-(16) of the main text. The static order parameter takes the value
\begin{equation}
  \label{eq:phi-ordered}
  \langle \phi \rangle = \pm\sqrt{\frac{-(r + r_\varepsilon)}{u}}
\end{equation}
where the sign is determined by initial conditions. Linearizing around the static order parameter value, $\phi(t) = \langle \phi \rangle + \varphi$ we find,
\begin{equation}
  \label{eq:nem-eom-ordered-fluc}
   \ddot{\varphi} +(\Gamma_{\rm nem} + \Gamma_{\varepsilon}) \dot{\varphi}  -2 (r + r_\varepsilon)\varphi  = 0\,,
\end{equation}
which describes stable oscillations around the order parameter value.

\begin{figure}
  \includegraphics[width=0.45\textwidth]{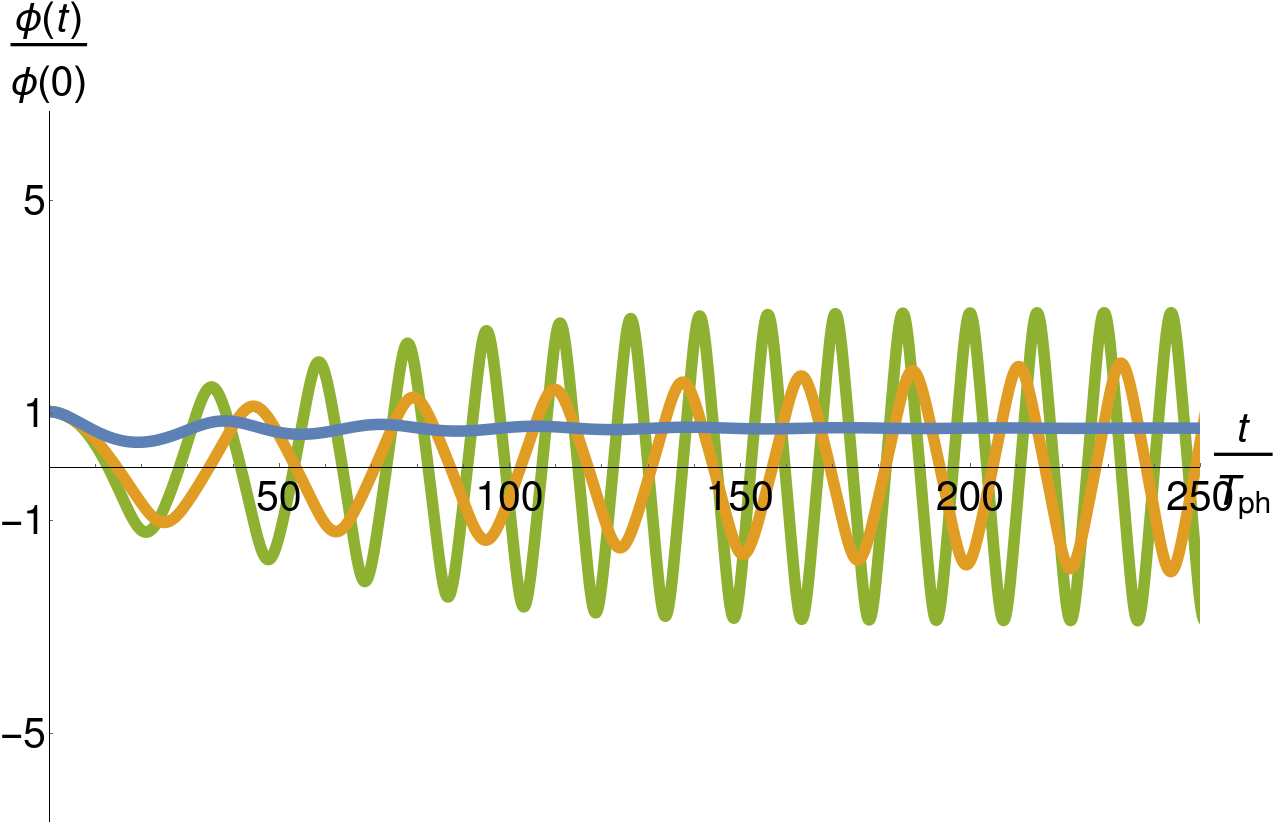}
  \caption{\label{fig:negative_r_small_damping_small_detuning}
    Classical evolution of the nematic order parameter
    for the case where $r<0$, and the laser is used to quench the nematic order. Here $\sqrt{-r}=0.025$, $\Gamma_{\rm ph} = 0.1$, and $\Gamma_{\rm nem}=0.01$. The laser is blue detuned from the phonon resonance and is set at $\Omega = 1.025$ (recall $\Omega_{\rm ph}=1$). The critical field is $\varepsilon_{0,c}\approx 0.120$. The color code is the same as in Fig. \ref{fig:small_r_small_detuning}.}
\end{figure}

\begin{figure}
\includegraphics[width=0.45\textwidth]{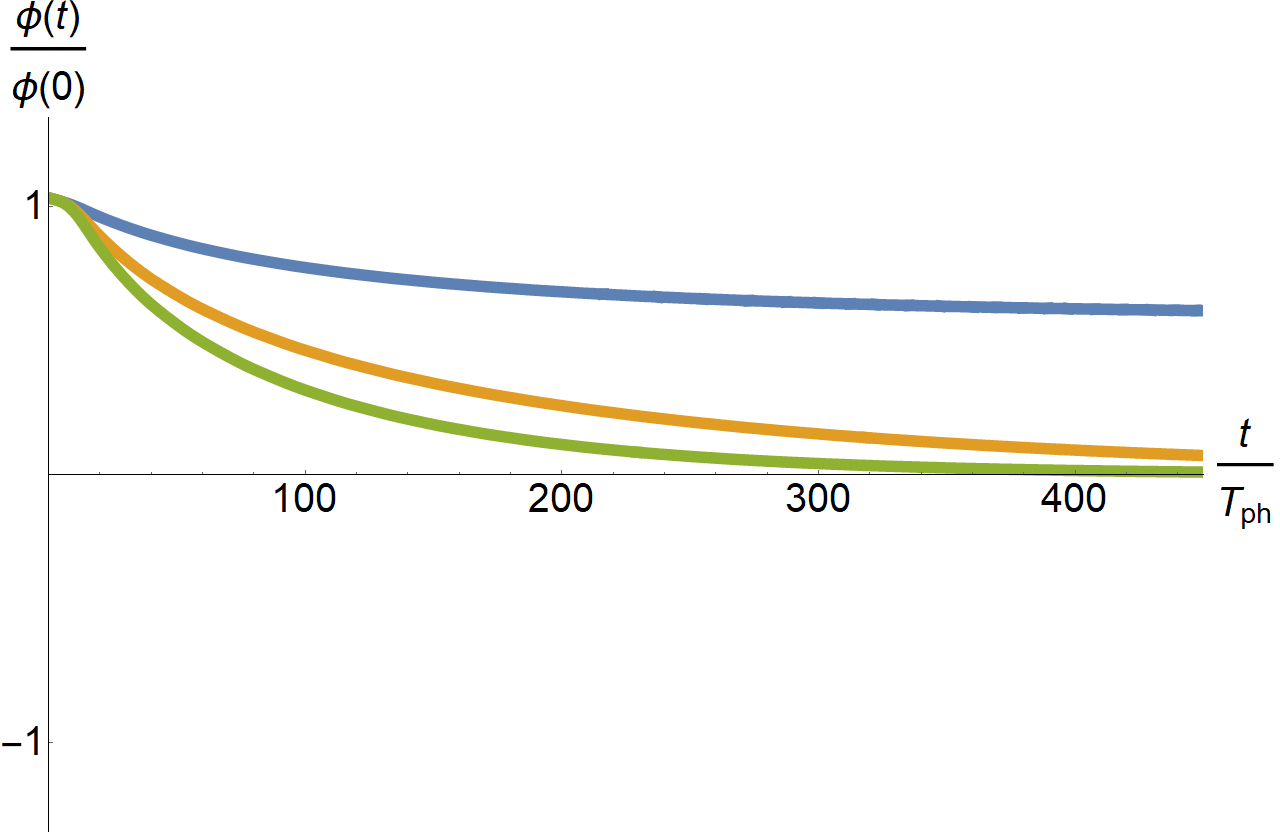}
\caption{
  Classical evolution of the nematic order parameter for the case where $r<0$, and the laser is used to quench the nematic order. In this specific case, $\sqrt{-r}\ll\Gamma_{\rm ph}$ and the Landau damping of the nematic propagator is assumed comparable to the phonon resonance frequency, $\Gamma_{\rm nem} \sim \Omega_{\rm ph}$. To be concrete, we choose $\sqrt{-r}=0.025$, $\Gamma_{\rm ph} = 0.1$, and $\Gamma_{\rm nem}=0.8$. The laser is blue detuned from the phonon resonance and is set at $\Omega = 1.0125$. The critical field is $\varepsilon_{0,c}\approx 0.145$. The color code is the same as in Fig. \ref{fig:small_r_small_detuning}.\label{fig:negative_large_r_small_damping_small_detuning}}
\end{figure}

\begin{figure}
  \includegraphics[width=0.45\textwidth]{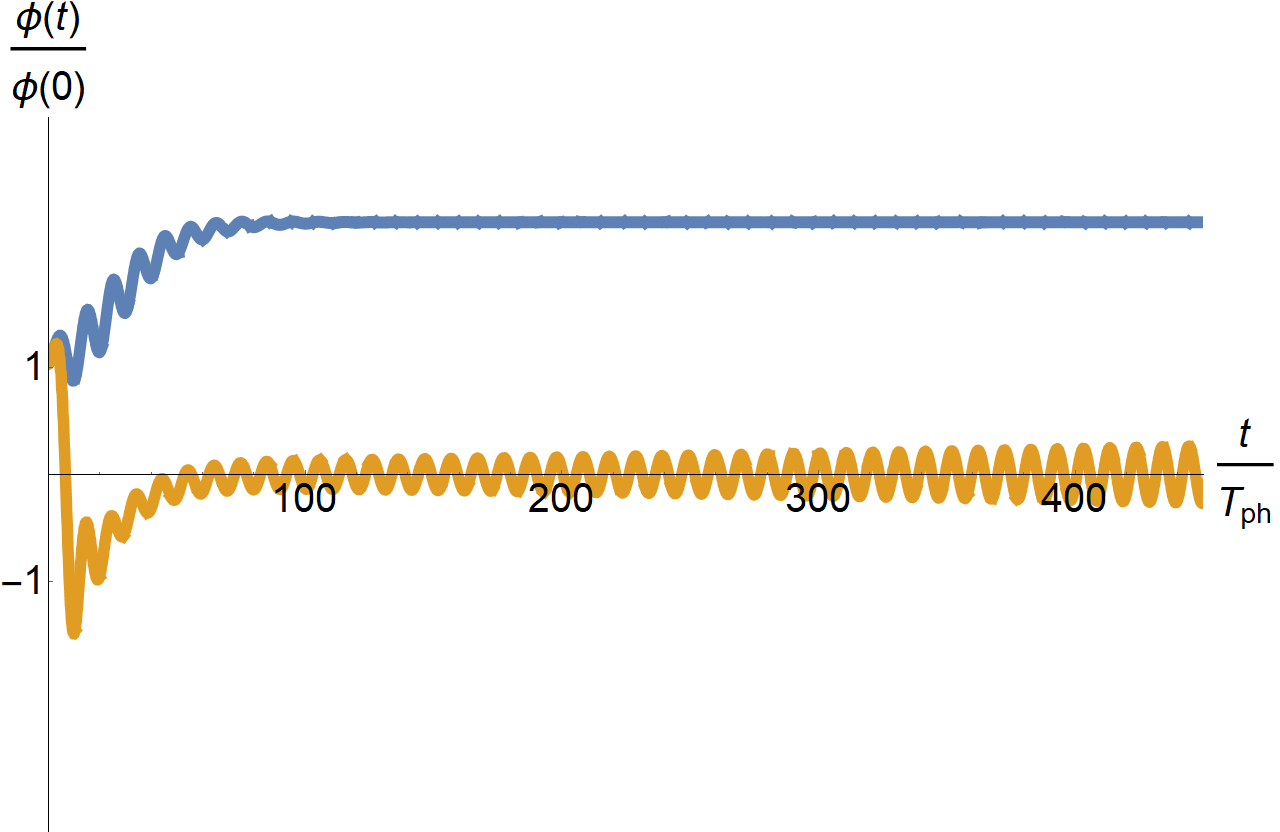}
  \caption{\label{fig:negative_large_r_large_damping_small_detuning}
    Classical evolution of the nematic order parameter for the case where $r<0$, and the laser is used to quench the nematic order. In this case, the Landau damping of the nematic propagator is assumed comparable to the phonon resonance frequency, $\Gamma_{\rm nem} \sim \Omega_{\rm ph}$, while the detuning, $\Omega - \Omega_{\rm ph} \gg \Gamma_{\rm ph}$ and $\sqrt{-r} \gg \Gamma_{\rm ph}$. Here, $\sqrt{-r}=0.25$, $\Gamma_{\rm ph} = 0.025$, and $\Gamma_{\rm nem}=0.8$. The laser is blue detuned from the phonon resonance and is set at $\Omega = 1.1$. The critical field is $\varepsilon_{0,c}\approx 0.979$. The color code is the same as in Fig. \ref{fig:small_r_small_detuning}; for clarity, the largest field (green curve) is omitted.}
\end{figure}

In figures \ref{fig:negative_r_small_damping_small_detuning}--\ref{fig:negative_large_r_large_damping_small_detuning}, we present solutions of the equations of motion for several initial conditions
with a
blue-detuned beam ($\Omega>\Omega_{\rm{ph}}$). In all cases, the system is driven out of the nematic phase when the electric field strength is larger than a critical value $\varepsilon_{0,c}$. Specific parameters for each case are displayed in the figure captions.
We do not present figures for the pole evolution for these parameter sets, since after the linearization in $\varphi$, all the pole information is already included in Eq. \eqref{eq:nem-eom-ordered-fluc}.

\subsection{The full nematic susceptibility}
\label{sec:full-effect-acti}

We now discuss the structure of the correction to the nematic susceptibility beyond the monochromatic laser approximation, and describe how we produced Fig.~3
of the main text. As we noted above, Eq.~\eqref{eq:chi-2-supp-time} gives the laser-induced correction to the quadratic part of the nematic action:
\begin{equation}
  \label{eq:S-nem-full}
  \delta S^{(2)} = \int \frac{d\omega_1 d\omega_2}{2\pi} \Phi(-\omega_1) \left[\chi^{-1}(\omega_1)\delta(\omega_1-\omega_2) + \delta \chi^{-1} (\omega_1,\omega_2)\right]\Phi(\omega_2),
\end{equation}
where
\begin{equation}
  \label{eq:chi-nem-def}
  \chi(\omega) =
  \begin{pmatrix}
     \chi^K(\omega) & \chi^R(\omega) \\
    \chi^A(\omega) & 0
  \end{pmatrix}, \quad
  \Phi(\omega) =  \begin{pmatrix}
  	\phi^{\rm cl} \\
  	\phi^{\rm q}
  \end{pmatrix},
\end{equation}
and the correction to the inverse nematic susceptibility
\begin{equation}
  \label{eq:chi-2-full-freq}
  \delta \chi^{-1} (\omega_1,\omega_2) = -\frac{2q^2\lambda^4}{a^4}\int \frac{d\nu}{(2\pi)^2}\varepsilon^{cl}(\omega_1-\nu)D^A(\nu-\omega_1)\begin{pmatrix}
                                           0 & D^A(\nu) \\
                                           D^R(\nu) & D^K(\nu)
                                         \end{pmatrix}  D^R(\nu-\omega_2) \varepsilon^{cl}(\nu-\omega_2)
\end{equation}
To understand the meaning of Eq.~\eqref{eq:chi-2-full-freq} it is convenient to consider the situation when $\delta \chi^{-1}(\omega_1,\omega_2) = F(\omega_1)\delta(\omega_1-\omega_2)$ is a function of just one frequency. It can be verified that this is the case when the laser beam has only one component,
$\epsilon^{cl}(t) \propto e^{-i\Omega t}$, such that the integrations in Eq. \eqref{eq:chi-2-full-freq} are constrained to $\omega_1=\omega_2,\nu -\omega_2 = \Omega$.
In that case the action, Eq.~\eqref{eq:S-nem-full}, depends only on one parameter, and furthermore obeys the fluctuation-dissipation theorem (FDT) at equilibrium,
\begin{equation}
  \label{eq:fluc-dis}
 \tilde{ \chi} ^K(\omega)  = \sigma(\omega)(\tilde{\chi}^R(\omega)-\tilde{\chi}^A(\omega)),
\end{equation}
where $\tilde{\chi} = (\chi^{-1} + \delta \chi^{-1})^{-1}$ is the renormalized nematic susceptibility, and $\sigma(\omega)$ is the sign function, which is the zero-temperature limit of the Bose-Einstein distribution function. Eq. \eqref{eq:fluc-dis} follows from the fact that both $\chi$ and $D$ obey the equilibrium FDT separately. The implication is that a pure monochromatic laser induces a quasi-equilibrium action for the nematic mode, provided that the nematic mode is irrelevant at frequencies of order $2\Omega$ and that superposition between positive and negative frequency elements can be neglected, as in e.g. a rotating-wave approximation. Recall that the electric field, being real, always has at least two frequencies, $\pm \Omega$, as implied by Eq. \eqref{eq:mono-action}. 

To study the effect of a light pulse which has a finite envelope, it is convenient to transform to the frequency difference and sum $\omega' = \w_1-\w_2$, $\omega=(\omega_1+\omega_2)/2$, so that,
\begin{equation}
  \label{eq:chi-2-full-freq-2}
  \delta \chi^{-1} (\omega,\omega') = -\frac{2q^2\lambda^4}{a^4}\int \frac{d\nu}{(2\pi)^2}\varepsilon^{cl}(\omega+\omega'/2-\nu)D^A(\nu-\omega-\omega'/2)\begin{pmatrix}
                                           0 & D^A(\nu) \\
                                           D^R(\nu) & D^K(\nu)
                                         \end{pmatrix}  D^R(\nu-\omega+\omega'/2) \varepsilon^{cl}(\nu-\omega+\omega'/2).
\end{equation}
If $\delta \chi^{-1} \propto\delta(\omega')$, the system is in quasi-equilibrium. For a slow enough pulse of length $\tau_{\rm pulse}$ much greater than the inherent timescales of the problem, dictated by $\Gamma_{\rm ph}$, $\Gamma_{\rm nem}$, etc., the $\omega'$-dependence is peaked near zero and can be dropped from the $\Phi$ fields. In that case we can transform $\delta \chi^{-1}$ from $\omega'$ back to time and obtain a quasistatic approximation for the nematic susceptibility correction $\delta \chi^{-1} (\omega,t)$.

\begin{figure}
  \centering
  \begin{minipage}{1.0\hsize}
    \includegraphics[width=0.5\hsize]{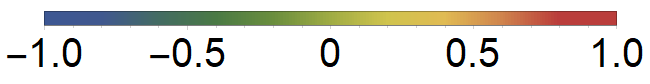}
  \end{minipage}
  \begin{minipage}{1.0\hsize}
    \includegraphics[width=0.23\hsize]{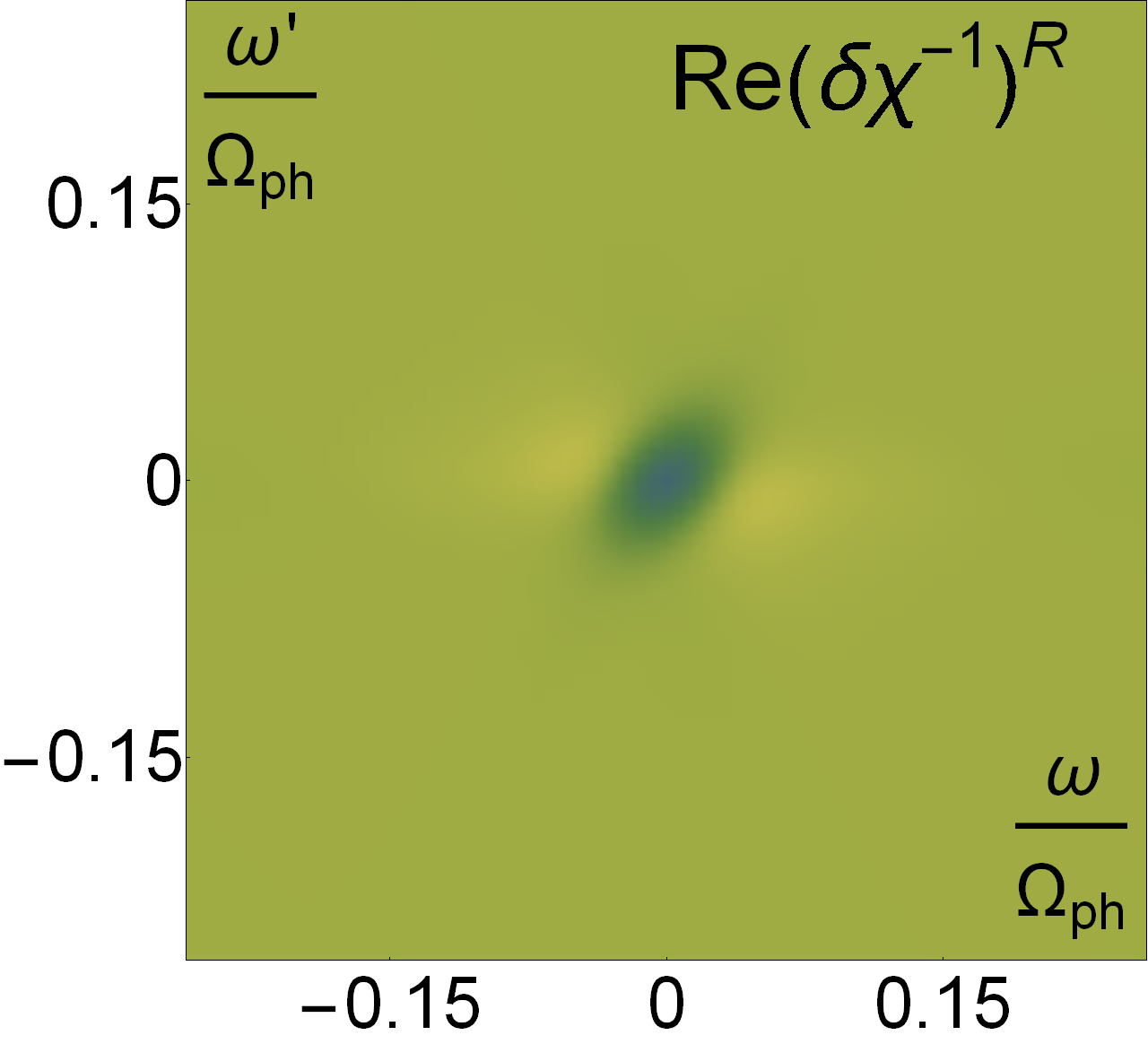}
    \includegraphics[width=0.23\hsize]{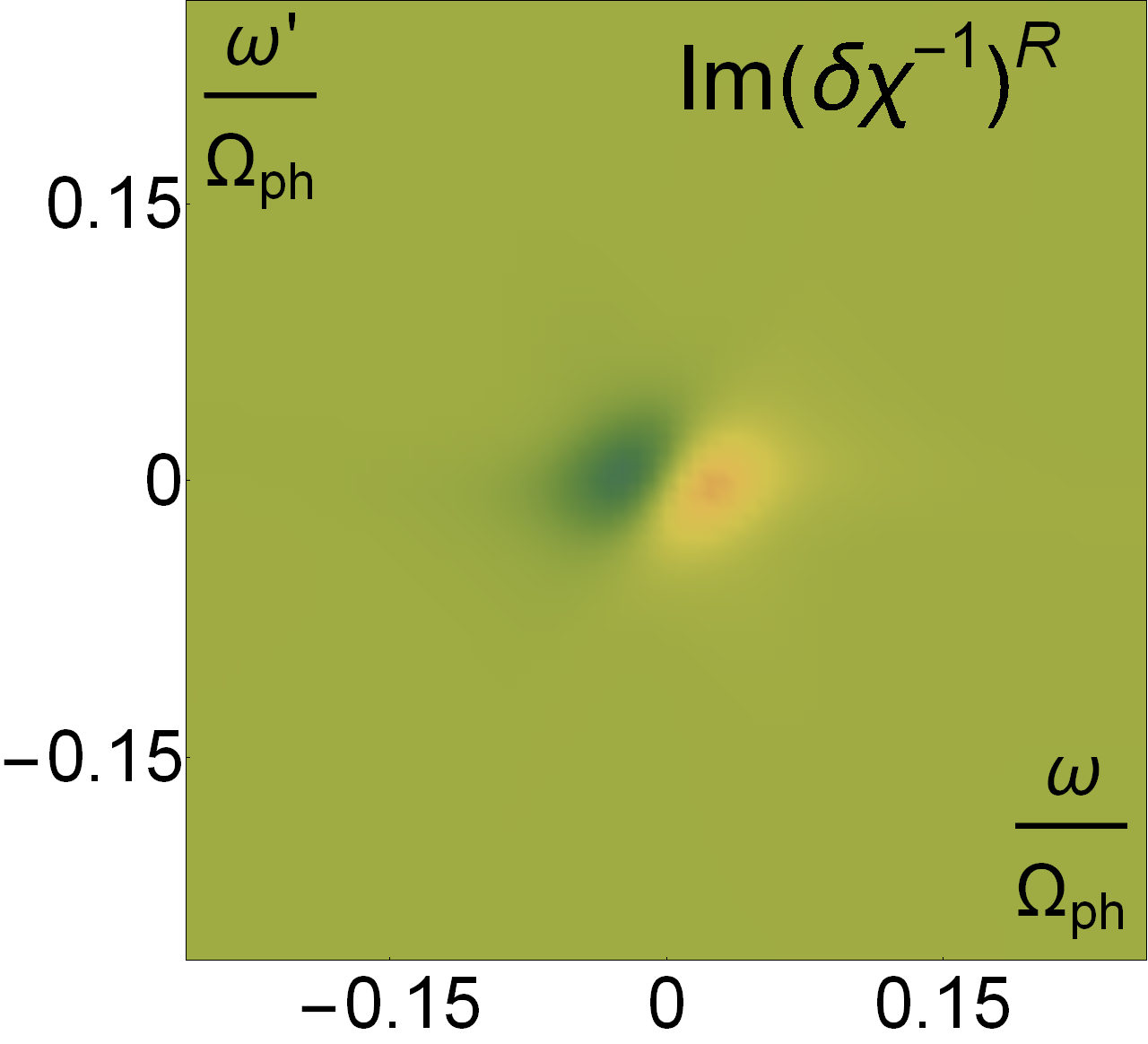}
    \includegraphics[width=0.23\hsize]{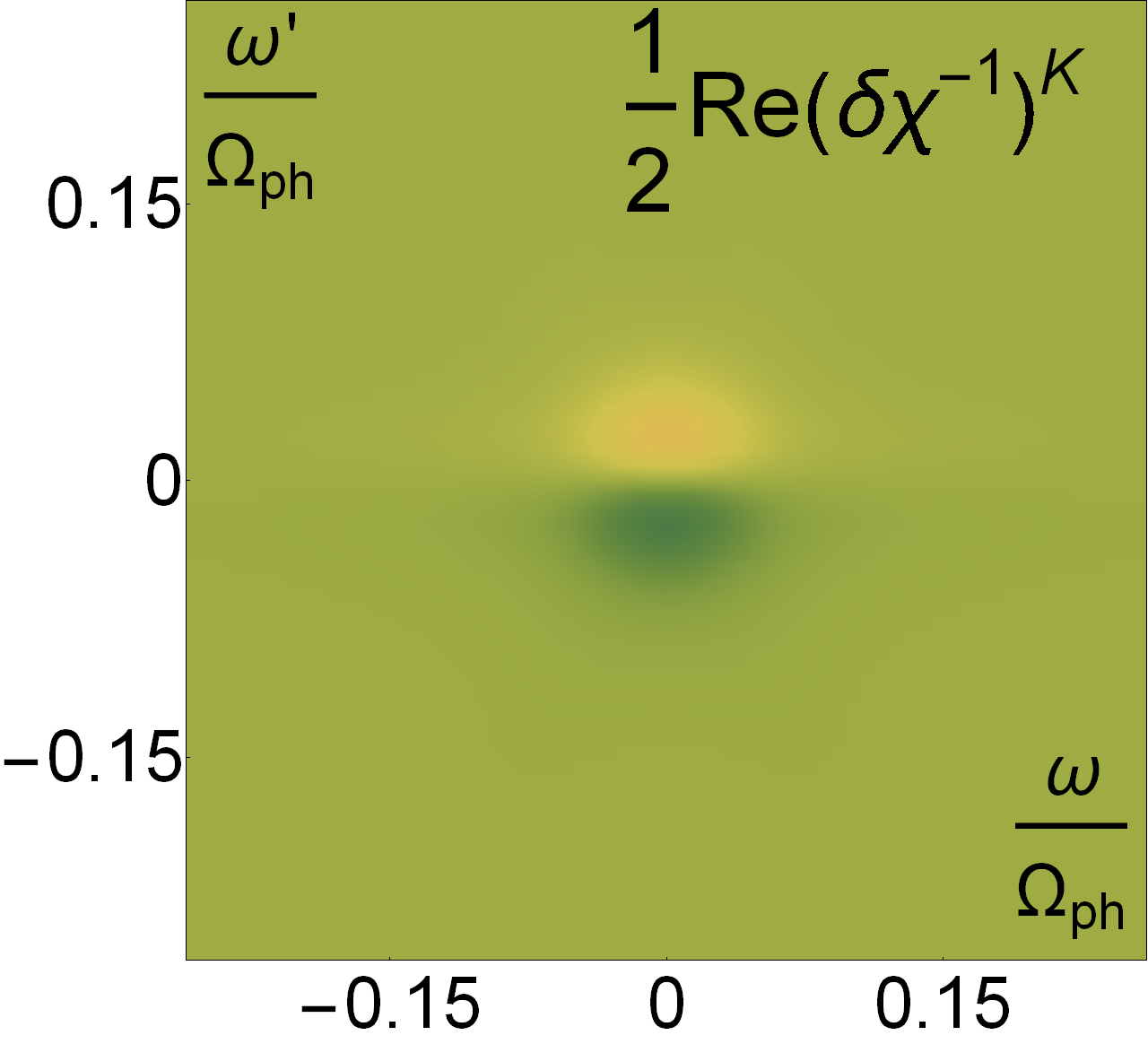}
    \includegraphics[width=0.23\hsize]{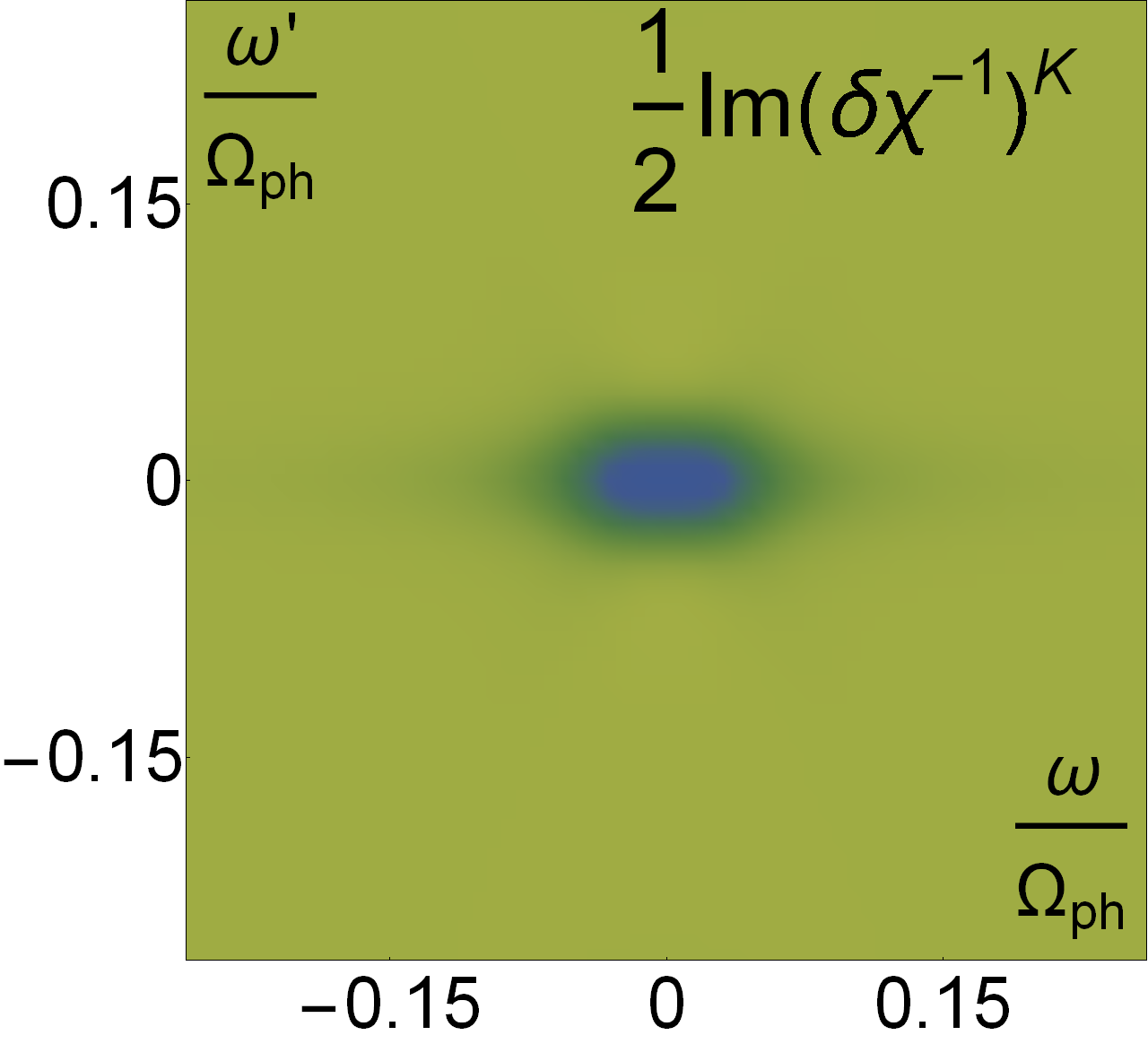}
  \end{minipage}
  \caption{The (normalized) retarded and Keldysh components of the correction to the nematic susceptibility $\delta (\chi^{-1})^{R/K}$ for a pulse at a frequency near the resonant phonon frequency, see Sec. \ref{sec:full-effect-acti} for parameter details. The Fourier transform in $\omega'-t$ space of the first two images are depicted in Fig. 3 of the main text. For clarity, we divided the Keldysh components by 2, since in equilibrium $|G^K| = 2|\text{Im}G^R|$.}
  \label{fig:induced-res}
\end{figure}

\begin{figure}
  \centering
  \begin{minipage}{1.0\hsize}
    \includegraphics[width=0.5\hsize]{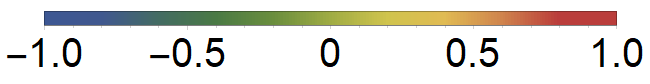}
  \end{minipage}
  \begin{minipage}{1.0\hsize}
    \includegraphics[width=0.23\hsize]{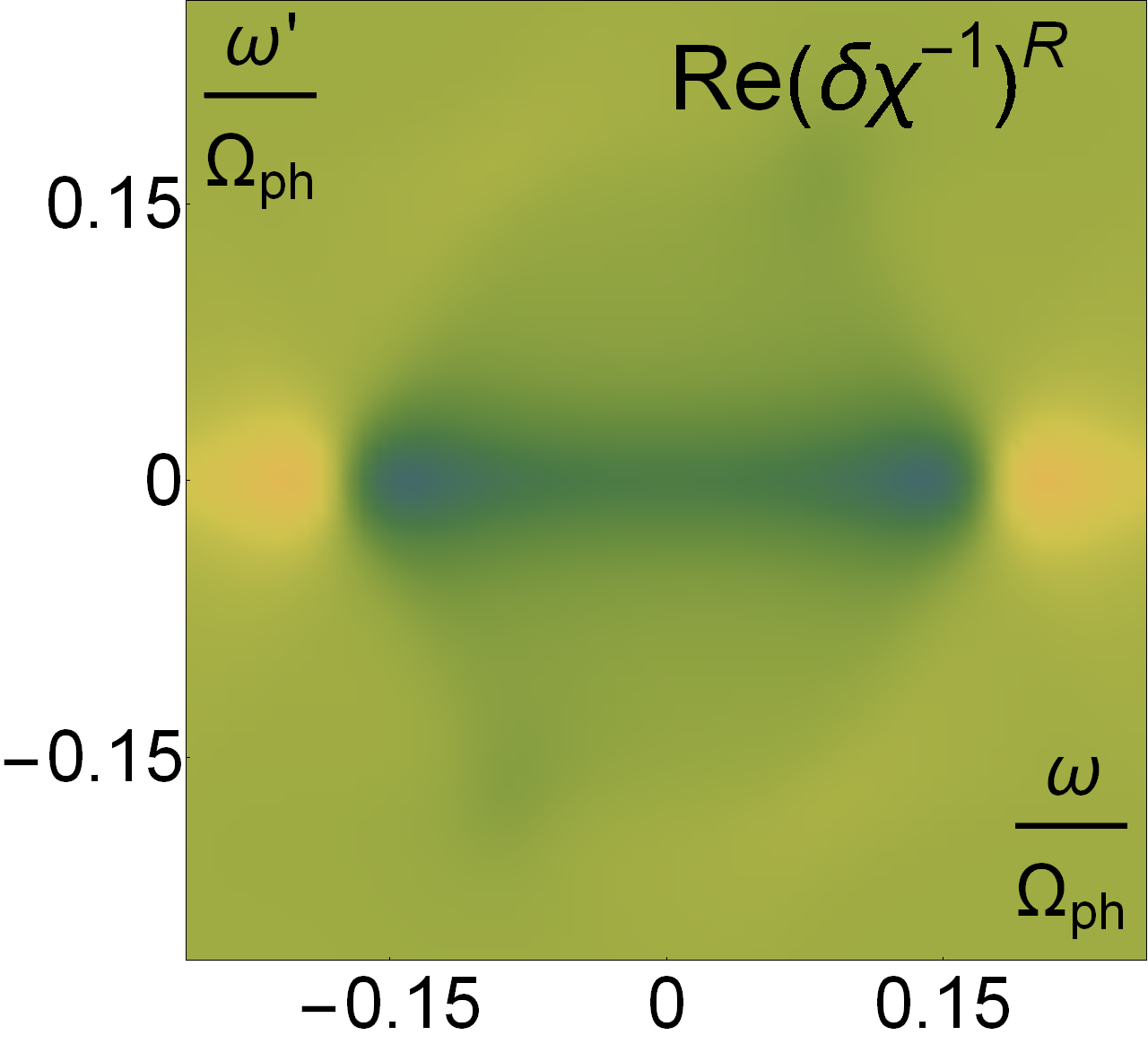}
    \includegraphics[width=0.23\hsize]{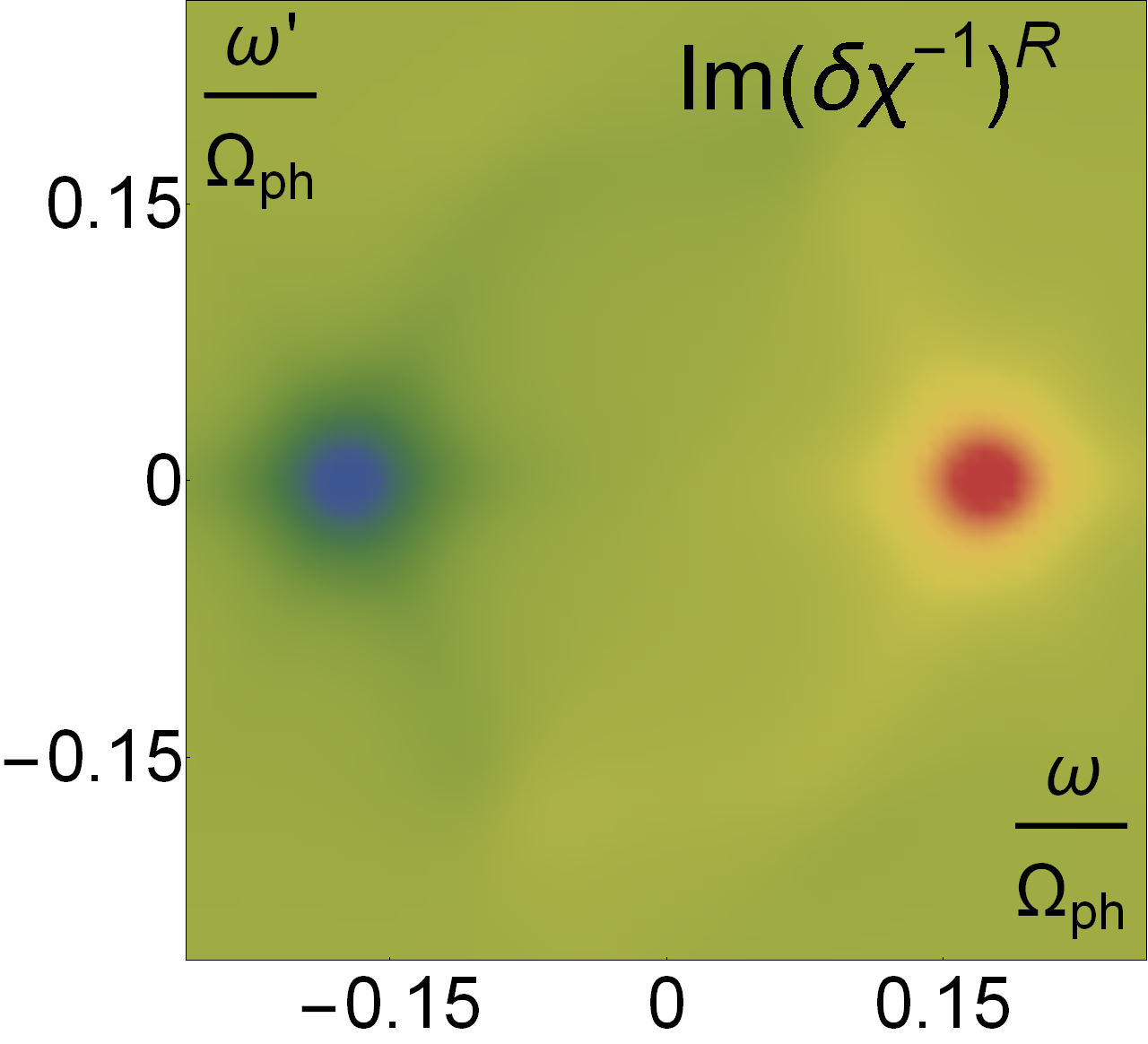}
    \includegraphics[width=0.23\hsize]{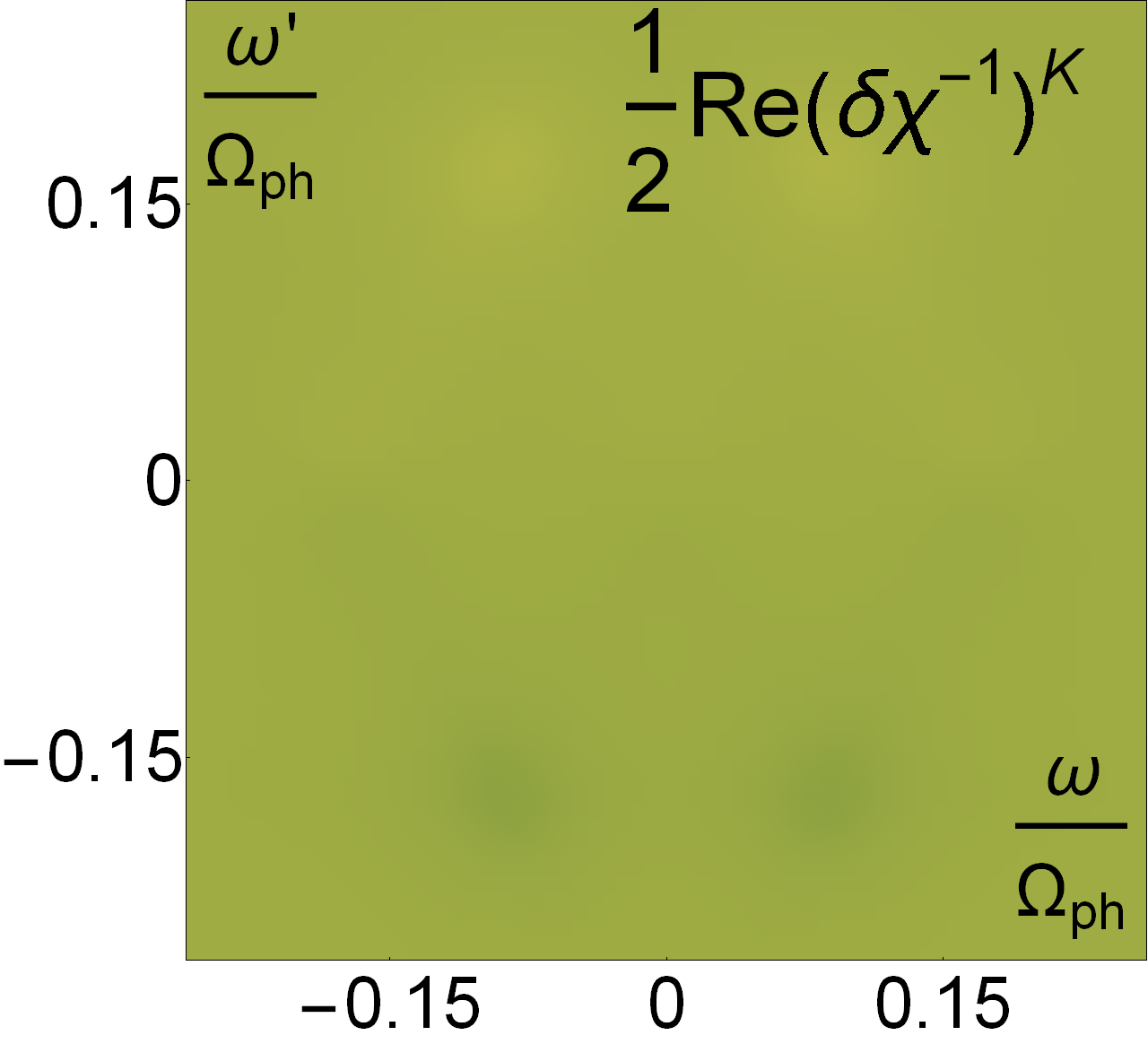}
    \includegraphics[width=0.23\hsize]{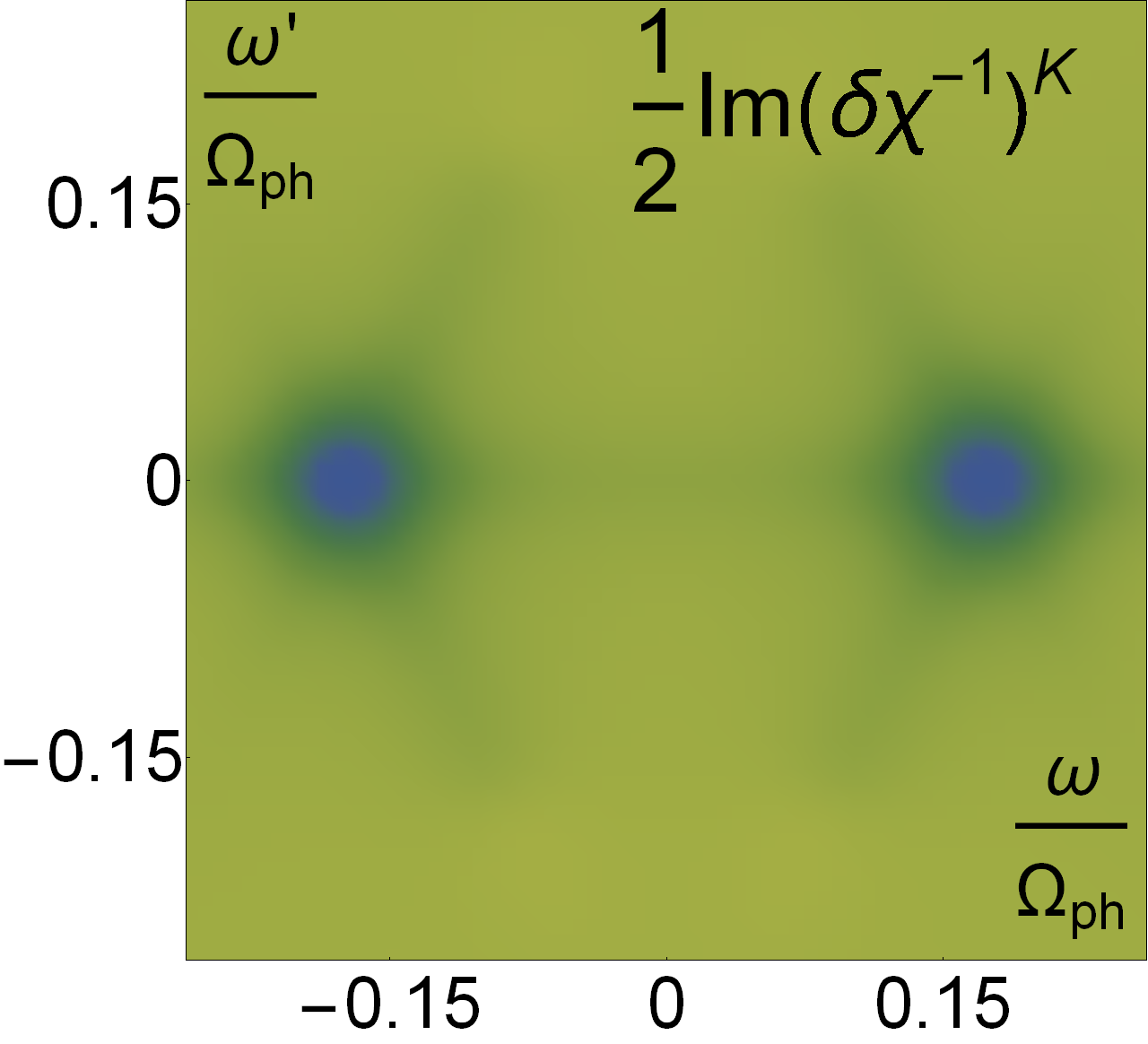}
  \end{minipage}
  \caption{The (normalized) retarded and Keldysh components of the correction to the nematic susceptibility $\delta (\chi^{-1})^{R/K}$ for a pulse at a frequency away from the resonant phonon frequency, see Sec. \ref{sec:full-effect-acti} for parameter details. The Fourier transform in $\omega'-t$ space of the first two plots are depicted in Fig. 3 of the main text. For clarity, we divided the Keldysh components by 2, since in equilibrium $|G^K| = 2|\text{Im}G^R|$.}
  \label{fig:induced-nonres}
\end{figure}

In Fig. \ref{fig:induced-res} we depict numerical computations of $\delta \chi^{-1}(\omega, \omega')$
for a pulse of the form
\begin{equation}
  \label{eq:pulse-laser}
  \varepsilon(t) = \frac{\hat x + \hat y}{\sqrt{2}}\varepsilon_0 \cos \left( \Omega t \right) \, e^{-|t|/\tau_{\rm pulse}},
\end{equation}
where we used the following parameters (normalized in the same way as in the previous section): $\Gamma_{\rm ph} = 0.05$, $\tau_{\rm pulse} = 6\pi/\Gamma_{\rm ph}$. In the main text,
Fig. 3(a) is a plot of $(\delta \chi^{-1})^R(\omega,t)$ obtained by a numerical Fourier transform with respect to $\omega'$. To obtain clear-looking images, we tuned the pulse near resonance, $ \Omega = 1- \Gamma_{\rm ph}/2$. Aside from the retarded component, in Fig.~\ref{fig:induced-res} we also present the Keldysh component of $\delta \chi^{-1}$. As Fig.~3 of the main text and Fig.~\ref{fig:induced-res} show, the quasistatic approximation is indeed a good one for these parameters.
Indeed, in Fig. \ref{fig:induced-res} the width of $\delta\chi^{-1}$ is narrower on the $\w'$ axis than on the $\w$ axis, implying that $\delta\chi^{-1}(\omega,t)$ evolves slowly in time for those frequency scales in $\w$ where $\delta\chi^{-1}$ is significant.
In order to calculate the classical response to the pulse, we solved Eqs.~\eqref{eq:phonon_eom}--\eqref{eq:nem_eom} for the same pulse, with the values of $\lambda = 0.04$, $\varepsilon_0 = 0.04$. We verified that the result does not change qualitatively for larger detuning. Indeed, in Fig. \ref{fig:induced-nonres} we depict the correction to the nematic susceptibility for an off-resonant pulse with $ \Omega = 1- 7\Gamma_{\rm ph}/2$. It can be seen that the quasistatic approximation is still a good one, and furthermore the correction to the nematic susceptibility has a quasi-equilibrium form. This follows from the fact that the induced part $\delta\chi^{-1}$ shows good agreement with the equilibrium FDT, see Eq. \eqref{eq:fluc-dis} and the discussion surrounding it. To see the good agreement, recall that by construction, $(\delta\chi^{-1})^{cl,cl} = 0$, while from the figure $(\delta\chi^{-1})^K \approx -2i\sigma(\omega)\text{Im}(\delta\chi^{-1})^R$, where $\sigma(x)$ is the sign function. Inverting the matrix yields $\delta\chi^K \approx 2i\sigma(\omega)\text{Im}(\delta\chi)^R$. 
As noted in the main text, for the on-resonant pulse depicted in Fig. \ref{fig:induced-res}, the nematic order-parameter is not in quasi-equilibrium, as can be seen by e.g. the significant induced real part of $(\delta\chi^{-1})^K$.

\subsection{Details of the numerical solutions}
\label{sec:deta-numer-solut}

The positions of the poles depicted in Fig.~2 of the main text and Figs. \ref{fig:small_r_small_detuning}--\ref{fig:negative_large_r_large_damping_small_detuning} in the supplementary material were obtained from analytical solution of Eq. \eqref{eq:poles_eq}. The classical trajectories in these figures and in Fig. 3 of the main text were obtained by numerical integration of Eqs. \eqref{eq:phonon_eom} and \eqref{eq:nem_eom}. The maps of $\delta \chi^{-1}$ in Fig. 3 were obtained by first numerically integrating the expression in Eq. \eqref{eq:chi-2-full-freq-2}, and convolving the result by a Gaussian mask that suppresses large frequencies, $|\omega_1,\omega_1| > 1/3$. This was done to remove the high frequency components of the field. This was then transformed back to the time domain by a numerical Fourier transform. The maps in Figs. \ref{fig:induced-res} and \ref{fig:induced-nonres} were obtained by numerically integrating the expression in Eq. \eqref{eq:chi-2-full-freq-2}.

 \section{Estimates of the experimental parameters for $\mathrm{FeSe}$}

\label{sec:appl-fe-superc-1}

In the main text, we invoked the iron-based superconductor FeSe$_{1-x}$S$_{x}$
to demonstrate the feasibility of our quantum quench protocol. In
this section, we briefly outline the sources and methods used to extract
experimental parameters for this compound. FeSe has been widely studied,
due at least partly to the fact that there are clean single crystals
available, and that it does not show long-range magnetic order at
ambient pressure~\cite{Boehmer2017}. Nevertheless, we were unable
to find experimental data for all the parameters used in our theory,
and therefore supplemented with data from other, related Fe-based
superconductors, most prominently FeTe and BaFe$_{2}$As$_{2}$.

Most of the external parameters in our theory are directly measureable
quantities. The exceptions are the coupling $\lambda$,
the maximum nematic order parameter strength $\langle\phi\rangle$,
and the bare nematic susceptibility $\chi_{0}/r$. In general,
the last two can only be extracted from experiment up to a prefactor
(which can also be temperature dependent). In a Stoner-type theory
for the nematic transition,
$\chi_{0}/r_0\sim N_{F}$, where $r_0$ is the bare nematic gap (at high temperatures) and
$N_{F}$ is the density of states at the Fermi level. 
Since the nematic susceptibility diverges at the nematic transition temperature $T_{\mathrm{nem}}$ with the mean-field behavior $(T-T_{\mathrm{nem}})^{-1}$, we estimate in these systems that $\chi_{0}^{-1}r=E_{F}(T-T_{\mathrm{nem}})/T_{\mathrm{nem}}$.
As mentioned in the main text, we took $\Delta k_{F}/k_{F}$
as a proxy for $\langle\phi\rangle$, which is again only correct
up to an unknown prefactor. We stress that the uncertainty in our
knowledge of $\chi_{0}^{-1}r$ and $\langle\phi\rangle$ means that
our estimates for the effective coupling are only valid to within
an order of magnitude.

For clarity, we have compiled our estimates of the various parameters
for FeSe into Table \ref{tab:exp-fe-based}. We now briefly outline
what sources we used to extract the experimental parameters in Table~\ref{tab:exp-fe-based}
and our estimate for the coupling, which, as noted in the paper, is
\begin{equation}
\lambda=\frac{\Delta\W}{2\langle\phi\rangle},\label{eq:lambda-estimate}
\end{equation}
where $\Delta\W$ is the splitting of the $E_{u}$ mode in the nematic
phase.

\begin{table}
\centering %
\begin{tabular}{|c|c|c|c|c|c|c|}
\hline 
$\hbar\Omega_{{\rm ph}}$  & $\hbar\Gamma_{{\rm ph}}$  & $\hbar\W_{{\rm nem}}$  & $\hbar\tau_{{\rm el}}^{-1}$  & $E_{F}$  & $k_{F}$  & $\Delta k_{F}$ \tabularnewline
\hline 
30-33  & 0.4 - 1.5  & 10  & 0.24  & 25  & 0.13  & 0.02 \tabularnewline
\hline 
\end{tabular}\caption{Estimated experimental parameters for the Fe-based superconductor
FeSe. Energies are quoted in ${\rm {meV}}$, and wavevectors in $\protect\AA^{-1}$.}
\label{tab:exp-fe-based} 
\end{table}

The infrared phonon structure and dispersion relations in FeSe have
been both calculated~\cite{Subedi2008,Wang2011,Wang2012} and detected
experimentally~\cite{Zakeri2017,Nakajima2017,Phelan2009,Ksenofontov2010}.
We discuss data only for the tetragonal phase, above $T_{\mathrm{nem}}\simeq90K$.
Ref.~\onlinecite{Nakajima2017} reported $\hbar\W_{{\rm ph}}=30.9~{\rm {meV}}$
in a film of FeSe on CaF$_{2}$, measured by optical reflectometry.
Ref.~\onlinecite{Zakeri2017} reported $\hbar\W_{{\rm ph}}=32~{\rm {meV}}$
near the $\bar{M}$ point, measured by electron energy-loss spectroscopy
on a single crystal. Ref.~\onlinecite{Ksenofontov2010} reported
$\hbar\W_{{\rm ph}}=31.3~{\rm {meV}}$ in neutron scattering. This tallies
with theoretical calculations~\cite{Subedi2008,Wang2012} predicting
$\hbar\W_{{\rm ph}}=30-35~{\rm {meV}}$, and only a weak dispersion for
the $E_{u}$ mode. We did not find a reported measurement of $\Gamma_{{\rm ph}}$
for FeSe. However, Ref.~\onlinecite{Homes2016} reported $\hbar\Gamma_{{\rm ph}}\sim1.2-1.5~{\rm {meV}}$
in the related chalcogenides FeTe and FeTe$_{1-x}$Se$_{x}$. For FeSe $B_{1g}$ optical mode, Ref.~\onlinecite{Gnezdilov2013}
reported a decay rate $\hbar\Gamma_{B_{1g}}\approx0.4~{\rm {meV}}$ at
the nematic transition.

The electronic structure and dynamics of FeSe have been extensively
studied by (among others) ARPES and Raman techniques. To estimate
the relevant timescale $\Omega_{\mathrm{nem}}^{-1}$, we considered
reports of polarization-resolved Raman data measuring the dynamic
response of the nematic mode in FeSe$_{1-x}$S$_{x}$~\cite{Massat2016,Blumberg2017}.
These measurements show a wide damped peak centered around $25~{\rm {meV}}$,
and extending to about $50~{\rm {meV}}$ before beginning to decay.
Although there are interesting features in the entire region (for
details see e.g. Ref.~\onlinecite{Blumberg2017}), the sharpest features
show up at frequencies below about $\hbar\Omega_{\mathrm{nem}}\sim10-12~{\rm {meV}}$,
and soften as one approaches the nematic transition, giving us the
estimate for $\Omega_{\mathrm{nem}}$ in Table~\ref{tab:exp-fe-based}.
To estimate the coupling constant $\lambda$, we used Eq.~\eqref{eq:lambda-estimate}.
As a proxy for $\langle\phi\rangle$ we took the elliptical distortion
of the hole-like Fermi surface $\Delta\kf=k_{F,x}-k_{F,y}$ at the
$Z$ point. We extracted the values for $E_{F}=25~{\rm {meV}}$, $k_{F}=0.13$
$\AA^{-1}$ and $\Delta k_{F}=0.02$ $\AA^{-1}$ from Ref.~\onlinecite{Coldea2017}
(all at the $Z$ point). We have not found a detailed study of the
$E_{u}$ mode splitting in FeSe. However, the $E_{g}$ mode, which
is Raman active but has almost the same resonance frequency as $E_{u}$,
has been measured. Ref.~\onlinecite{Hu2016} reports a maximum split
of $\hbar\Delta\W=0.4~{\rm {meV}}$ at $20~{\rm {K}}$. These estimates yield $\lambda \approx 1.3$~meV.

Using these numbers, we obtain an estimate for the shift of the nematic susceptibility. From Eq. (15) of the main text, the maximum shift occurs for
\begin{equation}
  \label{eq:shift-max}
  \Omega^2 = \W_{\rm ph}^2
   (1-\gamma)
\end{equation}
where $\gamma = \Gamma_{\rm ph}/\Omega_{\rm ph}$. The shift in the nematic transition temperature $T_{\rm nem}$ is proportional to the shift in $r$. For small $\gamma$, the shift is
\begin{equation}
  \label{eq:delta-r-new}
  \frac{\delta T_{\rm nem}}{T_{\rm nem}} = \frac{- r_\varepsilon}{\chi_0 E_F} \approx \frac{2\lambda^2}{E_F\hbar\Omega_{\rm ph}}n_{\rm ph}\times \frac{\W_{\rm ph}}{2\Gamma_{\rm ph}} \approx 0.045 n_{\rm ph}.
\end{equation}
Since $T_{\rm nem} \approx 90 $K, this corresponds to about $4$K per phonon.
To estimate the maximum possible phonon occupation number,
it is simplest to consider what occupation number would melt the lattice.
This can be found from the Lindemann criterion, 
\begin{equation}
n_{{\rm ph}}^L a^{2}=c_L^{2}\ell^{2},\label{eq:lindemann}
\end{equation}
where $\ell=3.7\AA$ is the $a$-axis lattice constant of FeSe
and $c_L$ is some fraction (we choose the commonly used value $c_L = 0.15$ ~\cite{Nelson2002defects}) and $a$ is the classical oscillator length, $\sqrt{\hbar/M\Omega_{\rm ph}}$.
Since the $E_{u}$ mode involves motion of
both the Fe and Se atoms, we use the mass, $M=\sqrt{M_{{\rm Fe}}M_{{\rm Se}}}\approx67u$. 
Then we find $a = 0.045\AA$ , which in turn implies $n_{\rm ph}^L = 152$.
A conservative estimate for $n_{\rm ph}$ is $n_{\rm ph} = 0.1 - 0.2 n_{\rm ph}^L$. To be concrete we take $n_{\rm ph}=0.15n_{\rm ph}^L$ resulting in $\delta T_{\rm nem} \approx 90 $K.

To estimate the equilibration time $\tau_{{\rm el}}$, we summed up
the phonon decay time $\Gamma_{{\rm ph}}^{-1}$ with measured electronic
decay times $\Gamma_{{\rm el-ph}}^{-1}$ from ultrafast optical reflectivity
experiments. Typically, such measurements heat up the electronic subsystem,
which then decays slowly into the lattice~\cite{Luo2012,Luo2012a}.
This decay is characterized by two distinct timescales: a fast decay
of the electrons into symmetry-preferred optical phonon modes (e.g.
$A_{1g}$) and then a slow anharmoic decay of these modes to the lattice.
We took as our estimate for the decay time $\Gamma_{{\rm el-ph}}^{-1}$
the decay constant of this slow anharmonic decay. We also considered
temperatures not too close to the critical temperature $T_{\mathrm{nem}}$,
out of the assumption that such a timescale roughly characterizes
a generic electron-phonon decay. We obtained $\Gamma_{{\rm el-ph}}^{-1}=0.5-0.6~{\rm {meV}}$
from Ref.~\onlinecite{Luo2012}. We took as our estimate for $\Gamma_{{\rm ph}}$
the lower value quoted above in Table \ref{tab:exp-fe-based} (which
is one that was measured for actual FeSe), implying a total decay
$\tau_{{\rm el}}=\Gamma_{{\rm ph}}^{-1}+\Gamma_{{\rm el-ph}}^{-1}=18.8-17.2~{\rm {ps}}$.
However, this estimate does not take into account the expected slowing
down of electronic heating rates near the nematic transition. Such
slowing down has been measured in BaFe$_{2}$As$_{2}$~\cite{Patz2014}.